\documentclass[12pt]{article}
\usepackage{graphicx}
\usepackage{setspace}
 \DeclareGraphicsExtensions{.eps, .ps}
\usepackage{soul,color}
\usepackage{xcolor}
\usepackage{amsmath}
\usepackage{amsthm}
\usepackage{amsfonts}
\usepackage{newfloat}
\usepackage{bbm}
\usepackage{natbib}
\usepackage{hyperref}

\usepackage{subcaption}
\usepackage[T1]{fontenc}

\usepackage{pifont}

\usepackage{setspace}
\usepackage{geometry}

\usepackage{authblk}

\newcommand{\re}{\textcolor{black}}

\newcommand{\blind}{0}

\begin{document}

\def\spacingset#1{\renewcommand{\baselinestretch}%
{#1}\small\normalsize} \spacingset{1}


\if0\blind
{
\title{The Landscape of College-level Data Visualization Courses, and the Benefits of Incorporating Statistical Thinking}

\author{Zach Branson\thanks{Assistant Teaching Professor, Department of Statistics and Data Science, Carnegie Mellon University}, Monica Paz Parra\thanks{Master's student in the School of Information, University of California, Berkeley}, Ronald Yurko$^*$}

\maketitle
} \fi

\if1\blind
{
  \bigskip
  \bigskip
  \bigskip
  \begin{center}
    {\LARGE\bf The Landscape of College-level Data Visualization Courses, and the Benefits of Incorporating Statistical Thinking}
\end{center}
  \medskip
} \fi

\bigskip
\begin{abstract}
Data visualization is a core part of statistical practice and is ubiquitous in many fields. Although there are numerous books on data visualization, instructors in statistics and data science may be unsure how to teach data visualization, because it is such a broad discipline. To give guidance on teaching data visualization from a statistical perspective, we make two contributions. First, we conduct a survey of data visualization courses at top colleges and universities in the United States, in order to understand the landscape of data visualization courses. We find that most courses are not taught by statistics and data science departments and do not focus on statistical topics, especially those related to inference. Instead, most courses focus on visual storytelling, aesthetic design, dashboard design, and other topics specialized for other disciplines. Second, we outline three teaching principles for incorporating statistical inference in data visualization courses, and provide several examples that demonstrate how to follow these principles. The dataset from our survey allows others to explore the diversity of data visualization courses, and our teaching principles give guidance for encouraging statistical thinking when teaching data visualization.
\end{abstract}

\noindent%
{\it Keywords:} course design, course survey, graphics, inference, statistics education
\vfill

\newpage
\spacingset{1.45} 

\section{Introduction: How is Data Visualization Taught, and What Defining Role Can Statistics Play?}

Nearly every field involves data visualization, where information from data are visualized in order to gain insight on some phenomena. Data visualizations can be as simple as a bar plot of categorical counts to as complex as an interactive dashboard involving maps, time series, networks, and other high-dimensional graphs. Given its ubiquity, data visualization is taught across many fields, each with their own unique goals. As a result, there are many books dedicated to data visualization within specific fields, including business \citep{soukup2002visual,knaflic2015storytelling,zheng2017data}, journalism \citep{cairo2012functional}, and the humanities \citep{engebretsen2020data}. There are also many books about data visualization in general \citep{tufte2001visual,few2004show,chen2007handbook,wilkinson2012grammar,wilke2019fundamentals,cairo2016truthful,cairo2019charts,schwabish2021better,cairo2023art,yau2024visualize,unwin2024getting} and data visualization software \citep{sarkar2008multivariate,jones2014communicating,wickham2010layered,wickham2016data,murray2017interactive,wickham2019welcome,sievert2020interactive}. Data visualization is such a pervasive and diverse area that, perhaps paradoxically, instructors in statistics, data science, and related fields may be unsure how to teach data visualization.

Indeed, the importance of data visualization in the practice and education of statistics has been emphasized for decades. For example, in the 2016 Guidelines for Assessment and Instruction in Statistics Education (GAISE) College Report endorsed by the American Statistical Association, one of the 10 goals is that students ``should be able to produce graphical displays and numerical summaries and interpret what graphs do and do not reveal'' \citep{carver2016guidelines}. \re{More generally, data visualization has been important for knowledge generation and communication for centuries (for historical reviews of data visualization, see, e.g., \citealt{tufte2001visual} and \citealt{friendly2008}). Maps for navigation were arguably the earliest form of data visualization; in the 1600s, visuals were used to understand variations in scientific measurements, which are often considered the first statistical visualizations; and in the 1800s, data visualizations were novelly used to argue for social and political change by, e.g., Florence Nightingale and W.E.B. Du Bois.} Although data visualization is likely used in most (if not all) introductory statistics courses, instructors and departments may wonder if there is value in teaching data visualization specifically, perhaps even in a standalone course. For example, our statistics and data science department at Carnegie Mellon University offers data visualization courses at the undergraduate and graduate level. And even if departments do not plan to dedicate a semester-long class to data visualization, they may wonder how they can incorporate instruction of data visualization into existing courses.

The purpose of this paper is to give guidance to instructors and departments in statistics and data science who may wish to incorporate data visualization into their curriculum. In particular, in this paper we try to answer two questions.
\begin{enumerate}
  \item[] \textit{Question 1}: How is data visualization currently taught across colleges and universities?
  \item[] \textit{Question 2}: What defining role can statistics play in teaching data visualization?
\end{enumerate}
To answer the first question, we collected data on 135 highly ranked colleges and universities, and determined which courses on data visualization, if any, were offered at each school in the 2022-2023 academic year. Among these 135 schools, we identified 270 data visualization courses. We recorded the name of each course, whether it was an undergraduate- and/or graduate-level course, the department(s) that offered it, and the topics and software taught based on its course description. There were three notable conclusions from our search.
First, many schools offered data visualization courses (94 out of 135, or 69.6\%), and often several departments within a school offered their own data visualization courses. The most frequent departments were related to computer science, data, information, and management, but we also found courses in less stereotypically data-centric fields, like geography, journalism, and the visual arts.
Second, we found that most data visualization courses were not taught in statistics and data science departments (220 out of 270, or 81.5\%), which may come as a surprise to some readers. \re{In part this is because we deemed a course as a data visualization course only if at least half of the course's material seemed dedicated to teaching data visualization itself. We found that while data visualization is frequently taught as one component of statistics courses (e.g., exploratory data analysis in introductory statistics courses), it is also rarely the focus of the course. Thus, introductory statistics courses were often not included in our survey.}
Finally, we found that most data visualization courses did not focus on statistical topics, and instead tended to specialize in visual storytelling, aesthetic design, or software use and development. In particular, the vast majority of courses did not focus on key aspects of statistical thinking: testing, uncertainty quantification, and modeling. For example, although many courses taught students how to visually examine trends and patterns in data, they rarely taught students how to visualize uncertainty in these trends or determine if these were ``statistically significant'' findings.

This brings us to our second question: What role can the field of statistics play in teaching data visualization? Data visualization has a long history in statistics: For example, John Tukey first introduced the boxplot in his book on exploratory data analysis \citep{tukey1977exploratory}, which was innovated upon with visuals like violin plots \citep{hintze1998violin} and sectioned density plots \citep{cohen2006sectioned} that display smoothed densities in addition to summary statistics. See \cite{wickham201140} for a history of the boxplot and its many extensions within the statistics literature. Statisticians have also been innovative in visualizing complex data structures, such as using trees and castles \citep{kleiner1981representing} or even faces \citep{chernoff1973use} to visualize high-dimensional data. Data visualization is also a routine part of any statistical practice. However, despite this history and practice, our survey suggests that recent college-level courses have not emphasized statistical aspects of data visualization. This does not necessarily mean that current courses are flawed: Our survey demonstrates that many high-quality data visualization courses exist. Rather, most of these courses are taught from a non-statistical perspective, and thus necessarily focus on the needs of other fields (e.g., storytelling in journalism, dashboard design in computer science, and specialized software for spatial data in geography and political science). Indeed, others have noted this disconnect between state-of-the-art data visualization and statistical principles in general \citep{gelman2013infovis,vanderplas2020testing}. This presents an opportunity for statistics-related departments to highlight the inferential nuances that arise when visualizing data, thereby taking more ownership in the enterprise of data visualization.

\re{To demonstrate how statistics departments can do this, we discuss how data visualization is taught in our statistics and data science department at Carnegie Mellon University, both in a 100-student semester-long undergraduate course (which serves many majors) as well as a 30-student half-semester master's course.} When our department first started teaching data visualization, it focused on visual design choices and how to use statistical software, which are common topics covered by many other college-level courses, as revealed by our survey. But over time, our course has evolved to put more emphasis on statistical issues that arise when creating data visualizations. For example, how should students pinpoint whether a graphical finding is ``statistically significant,'' and how can they quantify their uncertainty about trends and patterns they visualize? Furthermore, if multiple comparisons can be made within a single graph, how should the visual account for these multiple comparisons in order for inference to be correct? Addressing these questions forced us to develop a taxonomy of graphs and analyses, such that students can better understand how inferential tools can be properly used to complement their graphs, and vice versa. Our course isn't the only one that emphasizes statistical aspects of data visualization; in our survey, we found several other courses that emphasize similar topics. Nonetheless, these courses are by far the minority among data visualization courses.

We've found that there are two clear benefits to statistics-related departments teaching data visualization. First, it clarifies for students that creating data visualizations is often a statistical enterprise, such that it involves more than just coding skills and design choices. Indeed, if people wish to use visualizations to make a compelling argument and inform decision-making, then they will need to make inferential interpretations about those visualizations. Second, we've found that data visualization is a useful way to incorporate spiral learning \citep{harden1999spiral,bruner2009process} within our curriculum, where students revisit familiar topics (e.g., probabilities, distributions, hypothesis tests, statistical models) through the lens of visualization, which helps them understand these core topics at a deeper level.

The remainder of the paper is structured as follows. In Section \ref{s:survey} we describe our survey of college-level data visualization courses, and discuss our key findings. In Section \ref{s:inference} we outline three statistical principles for teaching data visualization: (1) graphs should be complemented with inference; (2) graphs are estimates and statistics that can be used in testing and uncertainty quantification frameworks; and (3) graphs can be used to motivate, teach, and interpret statistical analyses. To illustrate how instructors can use these principles to highlight inferential nuances in data visualization, we use examples from our courses at Carnegie Mellon University. In Section \ref{s:conclusion} we conclude with a discussion of further innovations that can be made based on our survey and teaching principles, as well as resources instructors can use to develop their own courses. In the supplementary material we give additional information about how our courses are structured for those who want to design their own course. Furthermore, we provide the data we collected from our survey of data visualization courses. We believe that this dataset will not only give further insights into how data visualization is currently taught, but can also serve as a useful dataset for students to explore when constructing their own visualizations.

\section{Survey Results for Data Visualization Courses} \label{s:survey}

\subsection{Description of Survey}

Similar to \cite{dogucu2022current}, who conducted a survey of undergraduate Bayesian statistics courses, we surveyed\footnote{We conducted this survey in June and July of 2023.} all universities with a U.S. News and World Report ranking of 100 or better and liberal arts colleges with a ranking of 50 or better \citep{us2023universities, us2023colleges, reiter2023data}. This consisted of 104 universities (due to ties) and 50 colleges. Among these universities and colleges, we identified 89 universities (85.6\%) and 46 colleges (92.0\%) whose online course catalogs were readily available and allowed textual search among courses offered.

For each university and college, we searched their online course catalog for credit-granting courses in the 2022-2023 academic year that contained one or more of the following words or combinations of words in the course title or description: (1) data visualization, (2) visualization / visualize / visualizing, (3) graph(s) / graphics / graphing, (4) plot(s) / plotting, and/or (5) chart(s) / charting. After identifying all courses that contained these words, we carefully reviewed the course titles, descriptions, and syllabi or websites (where available), and declared a course as a data visualization course only if its primary focus was indeed data visualization. \re{By ``primary focus,'' we mean that at least half of the course's material seemed dedicated to teaching data visualization itself. This criterion was meant to differentiate between courses whose primary motivation was teaching how to make effective data visualizations and courses that only used visualization in service of teaching a different concept. For instance, there are many introductory statistics and data science courses that cover data visualization in some capacity, e.g., using exploratory data analysis to describe a sample population, or using graphs to assess the plausibility of assumptions used for inference. However, in these classes, data visualization is usually only restricted to a small portion of the course, or visualization is only used to illustrate a different concept that was the main motivation for the course. Thus, we often did not consider introductory statistics courses as data visualization courses in our survey, unless their primary focus was indeed data visualization.} For simplicity, we excluded courses in professional programs (medical school, nursing school, law school, and MBA programs), such that we focus on general undergraduate and graduate programs. \re{Additionally, we only considered a course a data visualization course if its instruction on visualization was primarily data-centric, and not just visualization in general. For example, there are graphics courses that focus on creating video game graphics, which is a different context than visual representations of data that we are interested in. Similarly, there are many courses that focus on dashboard design or geospatial software, but only some of these involve data; we only considered these courses as data visualization courses if they indeed focused primarily on visualizing information from data (e.g., data dashboards or data-based maps).}

For each course that fit our search criteria, we gathered several pieces of information relevant to understanding the different contexts in which data visualization is taught. This includes the academic level (undergraduate, graduate, or both) and department in which the course was taught, as well as what software was used if listed in the course description or available syllabus/website. To additionally gauge the level of statistical content included in data visualization courses, we tracked if any of the following topics were mentioned in the course's description, syllabus, or website (if available): hypothesis testing, confidence intervals, statistical modeling (e.g., linear regression), spatial data, clustering, time series, high dimensional analyses (e.g., principal component analysis), text data, networks, and interactive graphics (e.g., Shiny, D3.js). \re{When possible, we also included a publicly available URL that was associated with each course, which others can use to inspect the content of a course. Course-specific URLs were available for 166 (61.5\%) of courses.}

All replication materials for our survey results, including the dataset of data visualization courses we identified in our survey, are available at \href{https://github.com/ryurko/teaching-data-viz}{\color{blue}github.com/ryurko/teaching-data-viz}. In what follows, we present several results about this dataset, thereby describing the current landscape of data visualization courses at U.S. colleges and universities. We hope that this dataset can also serve as a useful resource for instructors who want to glean further insights about data visualization courses, or use it as an example dataset in their own courses.

\subsection{Survey Results}

Out of the 135 universities and colleges with available online course catalogs, 94 schools (69.6\%) had at least one data visualization course according to our criteria. The majority of the universities ($78/89 = 87.6\%$) had at least one course, while fewer than half of the liberal arts colleges had at least one course ($16/46 = 34.8\%$).

\begin{figure}
    \centering
    \includegraphics[scale=0.7]{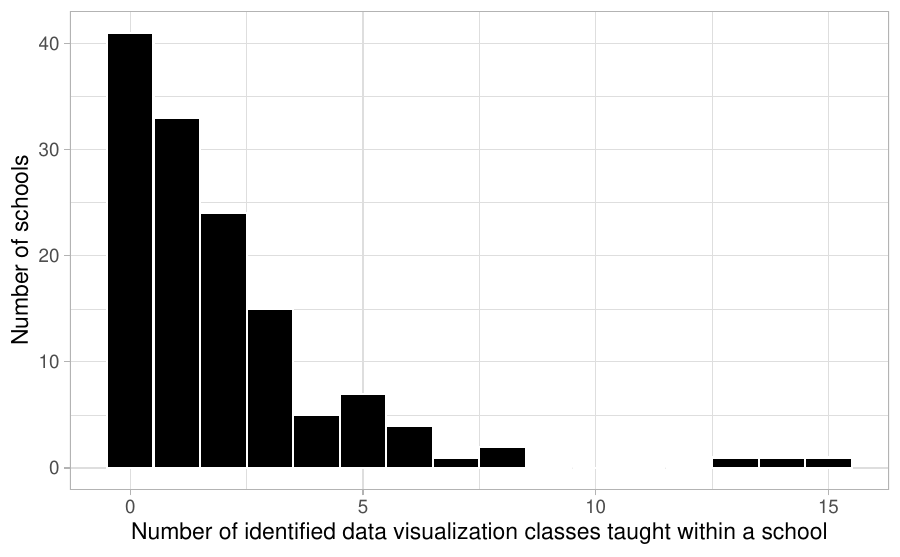}
    \caption{Number of data visualization classes taught within a school.}
    \label{fig:hist-n-course}
\end{figure}

In total, we identified 270 data visualization courses across this pool of universities/colleges. Figure \ref{fig:hist-n-course} displays the number of data visualization courses across schools; most schools have fewer than five courses, with a right-skew denoting places that offer several courses. Georgetown University (15 courses), New York University (14), and Northeastern University (13) stand out with the highest numbers. 

\begin{figure}
    \centering
    \includegraphics[scale=1]{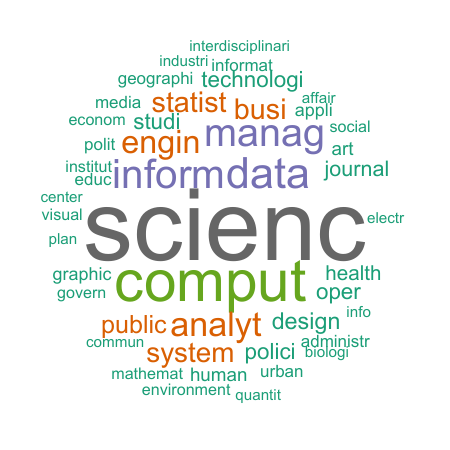}
    \caption{Word cloud displaying the top words in the department names for the 270 identified data visualization courses. All words displayed were used at least three times across department names.}
    \label{fig:dept-word-cloud}
\end{figure}

One of our main interests is understanding how many data visualization courses are taught by statistics and data science departments. Among the 270 courses, there are 163 unique department names, making it difficult to categorize courses by department. Indeed, departments ranged from quantitative fields (e.g., computer science, economics, engineering, mathematics, statistics) to social sciences (e.g., education, psychology, sociology) to professional domains (e.g., analytics, business, government, journalism) to humanities and the arts (e.g., anthropology, design, history, visual arts, writing), demonstrating the ubiquity of data visualization. Figure \ref{fig:dept-word-cloud} shows a word cloud of the most common words across department names among these 270 courses; \re{when creating this word cloud, we removed stop words (such that common words like ``of'' and ``the'' were removed) and performed stemming (such that words like ``science'' and ``sciences'' were combined as ``scienc''). Only words that were used at least three times across department names are displayed in Figure \ref{fig:dept-word-cloud}.} Broadly speaking, the most common departments are related to computer science, data, information, and management. We also see many departments related to analytics, business, engineering, public policy, and statistics. Finally, the diversity of less common departments is notable, including the arts, biology, geography, health, journalism, and mathematics.

\begin{figure}
    \centering
    \includegraphics[scale=0.7]{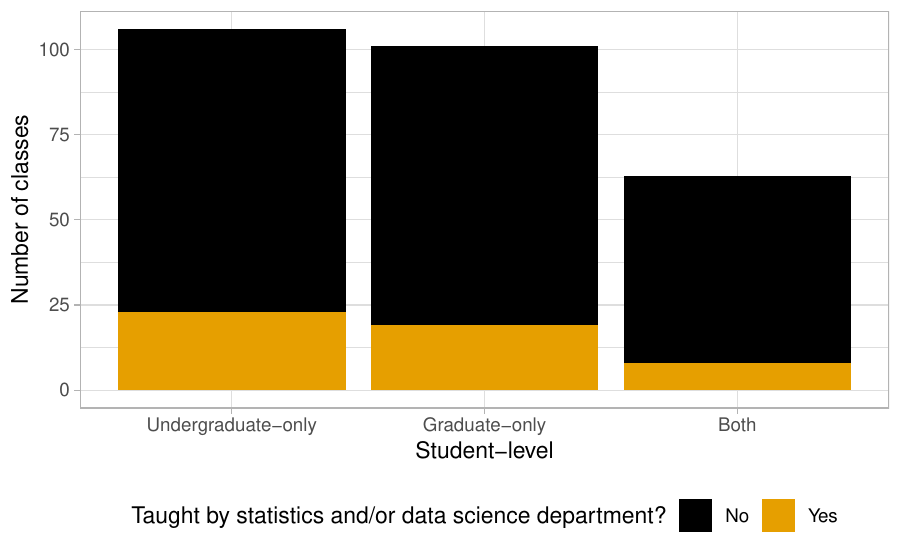}
    \caption{Number of data visualization courses that were taught by statistics and/or data science departments, sorted by student-level: undergraduate-only (21.7\%), graduate-only (18.8\%), or both (12.7\%).}
    \label{fig:dept-level-bars}
\end{figure}

To understand the involvement of statistics and/or data science departments specifically, we identified department names that included either ``stat'' or ``data'' in their name.\footnote{The only departments that mentioned ``data'' but not ``data science'' mentioned ``data analytics;'' the only exception was a department called ``Data for Political Research.'' In order to take an inclusive view of data science, we also considered these as data science departments.} We found 50 of the 270 courses (18.5\%) were taught by statistics and/or data science departments. This proportion was also fairly consistent across the level of courses: Figure \ref{fig:dept-level-bars} displays the number of data visualization courses at the undergraduate or graduate level, or both (e.g., cross-listed courses), colored by whether the course's department was in statistics or data science. A roughly equal number of data visualization courses are offered at the undergraduate or graduate level, where 21.7\% of undergraduate courses, 18.8\% of graduate courses, and 12.7\% of cross-listed courses are taught by statistics or data science departments. Taken together, Figures \ref{fig:dept-word-cloud} and \ref{fig:dept-level-bars} demonstrate that many fields teach data visualization, most of which are outside of statistics and data science.

This discrepancy in departments was also reflected in the topics that were taught in these courses. As mentioned in the previous subsection, we tracked whether courses taught various statistical topics (hypothesis testing, confidence intervals, modeling, high-dimensional analyses, clustering) and applications (spatial data, time series, text data, networks, and interactive graphics) by checking the course's description, website, or syllabus (if available). Rather than using exact text matches, we manually read each course's information and determined if each topic was indicated. For example, if a course's information mentioned linear regression, we considered this course as discussing statistical modeling, and if it mentioned principal component analysis, we considered this course as discussing high-dimensional analyses. Among the 270 identified data visualization courses, we were able to search for topics using either a description, website, or syllabus for 256 courses (94.8\%), such that the following results are reflective of the vast majority of courses we identified.

\begin{figure}
    \centering
    \begin{subfigure}[t]{0.44\textwidth}
    \centering
    \includegraphics[scale=0.55]{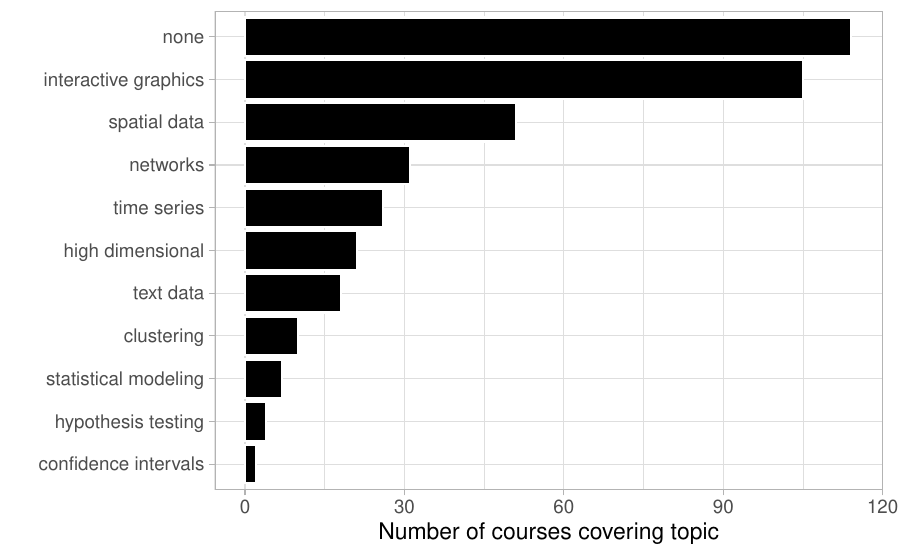}
    \caption{Number of courses that cover each topic. Note that ``none'' indicates that none of the 10 topics were taught in that course.}
    \label{fig:topic-count-bars}
    \end{subfigure}
    \hfill
    ~
    \begin{subfigure}[t]{0.44\textwidth}
    \centering
        \includegraphics[scale=0.55]{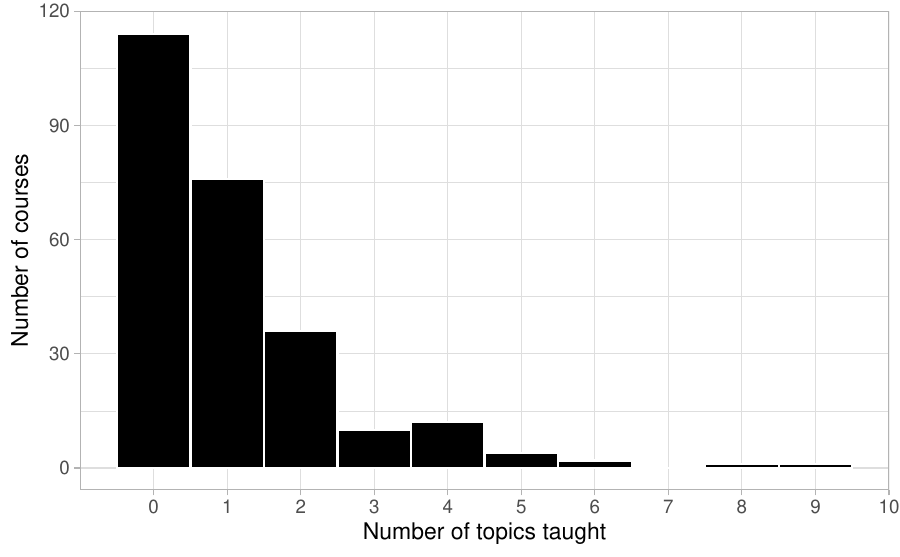}
        \caption{Number of topics taught across courses.}
        \label{fig:numTopics}
    \end{subfigure}
    \caption{Number of courses that cover each topic (left) and number of topics taught in each course (right) among the 256 courses for which we could identify topics, based on courses' descriptions, syllabi, or websites (if available).}
\end{figure}

Figure \ref{fig:topic-count-bars} displays the number of courses that covered each of the aforementioned topics. Out of the 256 courses for which we could ascertain topics, 114 (44.5\%) did not cover any of the listed topics. Instead, many courses appeared to focus on either ``storytelling'' (e.g., in journalism or business departments) or aesthetic design choices (e.g., in visual arts departments). Meanwhile, interactive graphics were by far the most popular topic: 105 (41.0\%) courses covered interactive graphics, whereas the second most popular topic was spatial data (51 courses, or 19.9\%). Interactive graphics was commonly taught in business, computer science, and information systems departments, whereas spatial data was commonly taught in geography, journalism, and political science departments. On the other hand, very few courses covered topics most related to statistical inference: Two courses covered confidence intervals (from Carnegie Mellon University and Colby College), four covered hypothesis testing (Brigham Young University, Carnegie Mellon University, Columbia University, and Duke University), and six covered modeling (Brown University, Carnegie Mellon University, Colby College, Duke University, Georgetown University, and University of Minnesota). Interestingly, not all of these courses were taught in statistics and/or data science departments; for example, the Brigham Young University course is in the political science department.

Meanwhile, Figure \ref{fig:numTopics} shows the number of topics that are taught across courses, and we see that the vast majority of courses (226, or 88.3\%) teach two or fewer of these 10 topics. For example, we found that many courses that taught interactive graphics focused exclusively on how to create data visualization dashboards, and many courses that taught spatial data focused exclusively on how to make data-based maps with geospatial software. The only course that covers nine topics is our own course in the Department of Statistics and Data Science at Carnegie Mellon University (which we discuss in the next section), and the only course that covers eight topics is from the Department of Statistics at Colby College.\footnote{The only topic our course doesn't cover is networks. Notably, the instructor who developed the Colby College course was previously a PhD student in our department and taught our data visualization course.} However, we want to emphasize that this \textit{does not} suggest that courses are inadequate by not covering these topics; many courses are certainly teaching other useful content, e.g., visual storytelling and design. Rather, departments understandably focus on topics most related to their own fields, instead of teaching statistical topics, especially those related to inference.

We acknowledge that there are limitations with this survey. \re{First, there are likely other data visualization courses not captured by our survey (e.g., at other schools not considered here, courses that did not mention keywords we searched for, or courses whose descriptions or syllabi are not reflective of the data visualization content taught in the course).} Similarly, some courses may cover statistical topics but did not mention them on publicly available sources we could access. Finally, this survey does not capture all possible topics taught in data visualization courses, because we focused on statistical topics. Nonetheless, this survey demonstrates that data visualization is taught across many colleges and universities, and by a broad range of disciplines that are largely outside of statistics or data science. Furthermore, most data visualization courses do not emphasize key aspects of statistical thinking (testing, uncertainty quantification, and modeling), and instead focus on topics unique to particular disciplines (e.g., storytelling, visual design, dashboard design, or specialized software). Thus, to complement current offerings in data visualization, in the following section we demonstrate how instructors can incorporate statistical thinking when teaching data visualization.

\section{Principles for Teaching Data Visualization to Encourage Statistical Thinking} \label{s:inference}

Our survey in Section \ref{s:survey} suggests that key aspects of statistical thinking---inference and modeling---are not emphasized in many data visualization courses. Thus, in this section we clarify how to teach data visualization in a way that encourages statistical thinking and highlights how inferential tools can enhance visualizations. In particular, we outline three statistical principles for teaching data visualization:

\begin{enumerate}
    \item \textbf{Graphs should be complemented with inference.} Using hypothesis tests and related inferential tools, users can determine whether what they see in a graph is a ``statistically significant'' finding. This is especially useful when using a graph to make an argument or aid decision-making.
    \item \textbf{Graphs are estimates and statistics.} Many graphs display summary statistics, but the graph itself is also a statistic with its own distribution. This view is especially useful for complex graphs, like maps, that may be a composite of many estimates.
    \item \textbf{Graphs can be used to motivate, teach, and interpret statistical analyses.} The first principle outlines how statistical methods can be used to better understand graphs; the opposite is also true. However, instead of trying to visualize a particular analysis, it is often fruitful to pinpoint an analysis that is most appropriate for a particular visualization. This clarifies the purpose of a given statistical method, and how to interpret its quantitative output.
\end{enumerate}

We \textit{do not} intend to argue that all data visualization courses should cover these principles. Rather, these principles suggest how to teach data visualization in a way that emphasizes statistical thinking, in conjunction with other skills (e.g., design choices and storytelling). To illustrate these principles, we use examples from our undergraduate course on data visualization in the Department of Statistics and Data Science at Carnegie Mellon University. First we briefly describe our course, so that instructors can understand how we scaffold the course to highlight these principles. Then, we discuss each principle, and demonstrate how they can be used to encourage statistical thinking alongside data visualizations.

\subsection{Description of Our Course, and a Taxonomy of Graphs}

Our undergraduate data visualization course in the Department of Statistics and Data Science at Carnegie Mellon University (CMU), titled Statistical Graphics and Visualization, has existed for almost 20 years and has evolved over time. Currently, the course is offered every fall and spring semester, and serves approximately 100 students each semester. The prerequisite is two semesters of introductory statistics (not calculus-based), such that students have a basic understanding of hypothesis testing, statistical models (including linear regression), and how to apply them to real datasets. We use R to implement graphs and analyses, but students are not required to have prior programming experience. \re{In addition to counting as an elective for our statistics and data science majors, the course also counts towards our college's general education requirement, specifically the ``design'' requirement, which consists of courses focused on designing products to solve problems. Our department is within CMU's Dietrich College of Humanities and Social Sciences, and we regularly teach majors in economics, English, history, information systems, psychology, social and decision sciences, and other fields, in addition to our own statistics and data science majors.} Thus, our course is meant for a broad group of students from many disciplines who are at a sophomore-level or above in terms of their statistical background.

\begin{table}[h]
    \centering
    {\tiny
    \begin{tabular}{l|l|l|l}
         \textbf{Variable Types} & \textbf{Fundamental Graphs} & \textbf{Statistical Quantities} & \textbf{Inferential Tools}  \\
         \hline
        1D Categorical & Bar plots, Spine charts & Counts, proportions & Chi-squared test for equal proportions \\
        \hline
        2D Categorical & \begin{tabular}{@{}l@{}} Stacked and side-by-side bar plots, \\ mosaic plots\end{tabular} & 
        \begin{tabular}{@{}l@{}} Contingency tables, proportions, \\ marginal and joint distributions \end{tabular} & \begin{tabular}{@{}l@{}} Chi-squared test for independence, \\ Pearson residuals \end{tabular} \\
        \hline
        1D Quantitative & \begin{tabular}{@{}l@{}} Histograms, smoothed densities, \\ empirical CDFs \end{tabular} &
        \begin{tabular}{@{}l@{}} mean, median, mode, \\ variance, skewness \end{tabular} & One-sample KS test \\
        \hline
        \begin{tabular}{@{}l@{}} 1D Quantitative  \& \\ 1D Categorical \end{tabular} & \begin{tabular}{@{}l@{}} Stacked histograms, \\ side-by-side boxplots, \\ overlayed densities \end{tabular} & \begin{tabular}{@{}l@{}} mean, median, mode, \\ variance, skewness, \\ marginal and conditional distributions \end{tabular} & \begin{tabular}{@{}l@{}} $t$-tests, one-way ANOVA, \\ Bartlett's test, two-sample KS test \end{tabular} \\ 
        \hline
        2D Quantitative & Scatterplots & \begin{tabular}{@{}l@{}} Mean, variance, \\ correlation, intercept/slope, \\ marginal and joint distributions \end{tabular} & Linear regression \\
        \hline
        \begin{tabular}{@{}l@{}} Above Scenarios \& \\ Additional Categorical Variables \end{tabular} & Facetted versions of above & Conditional distributions & \begin{tabular}{@{}l@{}} Subgroup analyses, \\ interactions in regression \end{tabular}
    \end{tabular}
    }
    \caption{Taxonomy of fundamental graphs, summary statistics, and inferential tools, based on different variable types. This constitutes the first-half of our semester-long undergraduate data visualization course.}
    \label{tab:taxonomy}
\end{table}

The first half of the course walks through a taxonomy of fundamental graphs, shown in Table \ref{tab:taxonomy}. The taxonomy is organized by the types of variables being considered, mainly combinations of one or two categorical and/or quantitative variables. The columns of Table \ref{tab:taxonomy}, from left to right, encapsulate our guide to students when making data visualizations: First consider which variables you want to visualize, then make graphs that are most appropriate for those variables, discuss statistical quantities relevant for those graphs, and use inferential tools to complement the graphs. In a typical week, one lecture is spent on the first three columns of Table \ref{tab:taxonomy}, where we discuss how to make and interpret graphs for that variable type, and then another lecture is spent on inferential nuances involved in interpreting those graphs. At the end of the week, students attend a computer lab where they practice making graphs and conducting analyses for a given data structure. We want to emphasize that Table \ref{tab:taxonomy} is not a comprehensive list of all topics that we cover in the first half of the course; for example, when discussing side-by-side boxplots and overlayed densities, we also discuss violin plots, ridgeline plots, and other graphical choices. Rather, Table \ref{tab:taxonomy} lists the prototypical graphs and analyses for a given data type, and illustrates how we organize graphical content alongside statistical content in a way that's easy for students to navigate.

\re{Some readers may wonder how the taxonomy in Table \ref{tab:taxonomy} is related to the grammar of graphics framework, which is a common conceptual model for organizing, implementing, and teaching data visualization \citep{wickham2010layered,wilkinson2012grammar}. In short, the grammar of graphics framework outlines seven components that map data to aspects of a graph; this includes how data are mapped to the geometry, coordinate system, and aesthetics of a graph. Table \ref{tab:taxonomy} includes two of these components: the variables and statistics that are used for a graphic. We view the grammar of graphics framework and the taxonomy in Table \ref{tab:taxonomy} as complementary to each other: The grammar of graphics framework formulates how graphs themselves are constructed, and Table \ref{tab:taxonomy} formulates how prototypical graphs are related to statistical inference. This way, students can more easily understand how graphs fit into a statistical inference paradigm.}

Because of this, we've found that there are several pedagogical benefits to organizing the first half of the course with the taxonomy in Table \ref{tab:taxonomy}. First, students are typically quite good at determining whether a variable is categorical or quantitative, but they may find it challenging to determine which graphs or analyses are most appropriate for a given dataset. Thus, organizing by variable type gives students a straightforward scaffolding for navigating graphs and analyses. Second, it prepares students to make and interpret more complex graphs in the second half of the course (e.g., high-dimensional visualizations, maps, interactive graphs), which frequently use the graphs in Table \ref{tab:taxonomy} as building blocks. Third, the taxonomy encourages a spiral learning approach in our undergraduate curriculum: Our students are often familiar with many of the graphs and analyses in Table \ref{tab:taxonomy}, and thus our course exposes them again to these topics. However, by covering these topics in tandem, students often gain a deeper understanding of inferential tools by seeing how they operate graphically. Finally, the taxonomy emphasizes that graphs and inferential tools can and should complement each other, which encourages the three statistical principles mentioned at the beginning of this section.

The second half of the course discusses more complex data structures: high-dimensional data, clustering, spatial and temporal data, text data, and interactive graphs. Although each data structure has its own unique challenges, many visualizations and analyses use those in Table \ref{tab:taxonomy} as building blocks. For example, students learn how to use principal component analysis and multi-dimensional scaling to visualize high-dimensional data with scatterplots, how to use regression to visualize kriging on maps for spatial data, and how to create animations by ``stitching'' these fundamental graphs together based on time or another variable. Thus, the statistical principles emphasized in the first half of the course naturally translate to the more complex visualizations in the second half of the course. The course culminates into a final project, where students make public-facing graphs and analyses using a real dataset. These graphs and analyses are hosted on a department website, such that students can reference their projects in applications to internships, jobs, and graduate programs. Thus, the course is structured to organically facilitate a data science portfolio for students \citep{nolan2021promise}, which helps them better see how the course material relates to their professional development. For instructors interested in developing their own data visualization course, we discuss further details about course design in the supplementary material.

In the remainder of this section, we discuss each of the three above statistical principles in depth, using examples from our course. These examples demonstrate how instructors can facilitate statistical thinking when teaching data visualization. Although we focus on examples from our undergraduate course, these statistical principles are applicable to other pedagogical levels and modalities. For example, we've also taught these principles in our master's program, as we discuss in the supplementary material. All replication materials for our examples can be found at \href{https://github.com/ryurko/teaching-data-viz}{\color{blue}github.com/ryurko/teaching-data-viz}.

\subsection{Principle 1: Graphs should be complemented with inference}

Many courses illustrate the importance of visualization by demonstrating how graphs can reveal phenomena that statistical analyses may hide. A classic example is Anscombe's quartet \citep{anscombe1973graphs}, which consists of four $(X,Y)$ datasets whose summary statistics are the same but whose scatterplots are quite different. Since \cite{anscombe1973graphs}, there have been many extensions of simulated datasets whose inferential results are identical but whose graphs differ (e.g., \cite{chatterjee2007generating,matejka2017same,healy2018data}). When teaching data visualization, we've found it useful to also illustrate the opposite phenomenon: graphical displays that are identical but whose inferential results differ. This demonstrates how statistical analyses can reveal phenomena that graphs may hide.

\begin{figure}[h!]
    \centering
    \includegraphics[scale=0.6]{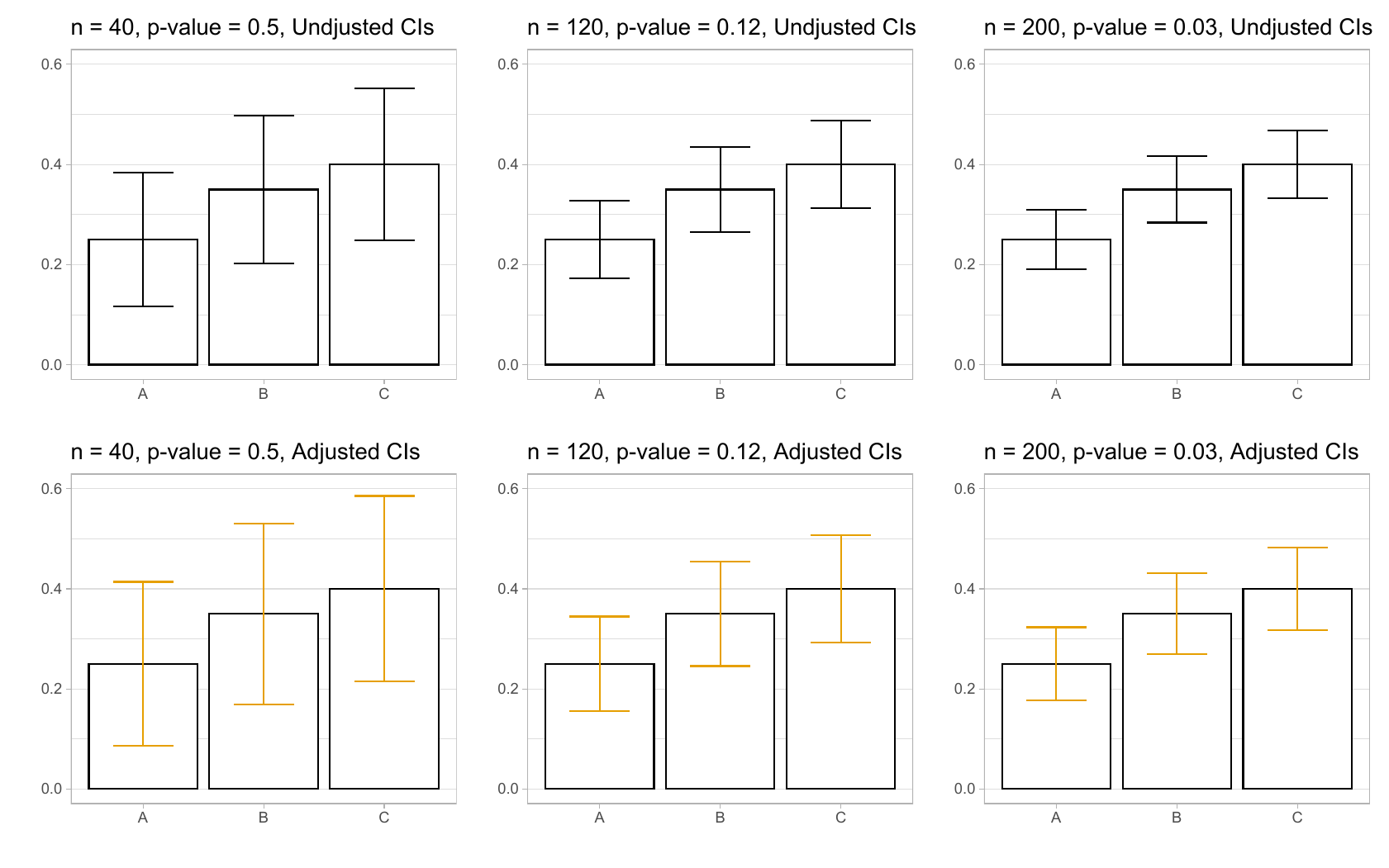}
    \caption{Bar plots displaying proportions $(0.25, 0.35, 0.4)$ for a single categorical variable; each y-axis is on the proportion scale. Plots in the left, middle, and right columns correspond to sample sizes $n = 40$, 120, and 200. Plots in the top row display 95\% confidence intervals (CIs, in black), and plots in the bottom row display Bonferroni-corrected CIs that account for all pairwise comparisons (in orange). The title of each plot also lists the $p$-value for a chi-squared test for equal proportions across the three categories.}
    \label{fig:barplots}
\end{figure}

Figure \ref{fig:barplots} shows a simple example using a bar plot to visualize one variable with categories ``A,'' ``B,'' and ``C.'' All six bar plots in Figure \ref{fig:barplots} display the same proportions $(0.25, 0.35, 0.4)$, but inferential conclusions about these plots differ, depending on the overall sample size $n$ (denoted by the columns of the figure) and whether or not we correct for multiple hypothesis testing (denoted by the rows of the figure). As a result, statistical analyses affect how we may interpret these graphs, despite them showing the same summary statistics. This introduces students to many inferential nuances that are useful to grapple with, as we discuss below.

First, one can conduct a chi-squared test to assess if the three proportions are equal; the test only rejects at the $\alpha = 0.05$ level for the $n = 200$ plots, and thus we may conclude that only these plots display evidence that the proportions are not all equal. Although many students know that sample size impacts hypothesis tests, many frequently do not consider how this can impact the interpretation of a graph. This also introduces the notion of statistical power for graphs, because sample size impacts the power of hypothesis tests. 

A natural follow-up question to the chi-squared test is: If the proportions are not all equal, which ones differ? To answer this question, we ask students to visualize confidence intervals for each proportion, and interpret the intervals accordingly. The top row of Figure \ref{fig:barplots} displays marginal 95\% confidence intervals for each proportion.\footnote{Specifically, for each estimated proportion $\hat{p}$ we computed the confidence interval as $\hat{p} \pm z_{\alpha/2} \sqrt{\frac{\hat{p}(1-\hat{p})}{n}}$, where $z_{\alpha/2}$ is the standard Normal quantile for $\alpha = 0.05$.} Frequently, students interpret a lack of overlap between intervals as a statistically significant difference (which is correct), but also interpret overlap as an insignificant difference (which is incorrect); see \cite{cumming2005inference} and \cite{wright2019primer} for detailed discussions of this misconception and how to address it in visualizations. For example, the ``A'' and ``C'' intervals slightly overlap in the $n = 120$ plot, and thus students are tempted to fail to reject the null hypothesis that these two proportions are equal. However, zero is not contained in the 95\% confidence interval for the difference in these proportions\footnote{A useful exercise is to ask students to compute the 95\% confidence interval for the difference in these two proportions, say $p_A - p_B$. If we view these two proportions as independent, then the confidence interval would be $\hat{p}_A - \hat{p}_B \pm z_{\alpha/2} \sqrt{ \frac{\hat{p}_A(1-\hat{p}_A)}{n} + \frac{\hat{p}_B(1-\hat{p}_B)}{n} }$. However, if we view the single categorical variable in Figure \ref{fig:barplots} as a multinomial random variable, then technically the standard error should be $\sqrt{ \frac{\hat{p}_A(1-\hat{p}_A)}{n} + \frac{\hat{p}_B(1-\hat{p}_B)}{n} + \frac{2 \hat{p}_A \hat{p}_B}{n} }$, in order to account for the covariance between categories. However, this does not affect the inferential conclusions we would make for the example in Figure \ref{fig:barplots}. }, and thus one should reject this null hypothesis. Interestingly, this occurs even though we fail to reject the null hypothesis that all proportions are equal using the chi-squared test. These findings---which are often surprising to students---demonstrate several statistical lessons, such as the difference between visual overlap of confidence intervals and statistical significance, the concept of power, and caution to not ``accept'' a null hypothesis. But most importantly, this exercise teaches students to be cognizant of null hypotheses they implicitly reject or fail to reject when interpreting a graph, and to be knowledgeable of how to formally test those hypotheses to complement the graph.

Indeed, when visually interpreting a graph, it's natural to make multiple comparisons, such as when we compare confidence intervals. This introduces a multiple testing problem, because we are visually making many comparisons within a single graph \citep{wright2019primer}. The bottom row of Figure \ref{fig:barplots} displays confidence intervals using a Bonferonni correction for all three paired comparisons; this is equivalent to displaying $100(1 - 0.05/3)\% = 98.3\%$ confidence intervals. This causes more intervals to overlap and changes some testing conclusions for the $n = 120$ plot, as shown in Table \ref{tab:testResults}, which summarizes inferential conclusions that arise from the six bar plots in Figure \ref{fig:barplots}. Walking through this example with students demonstrates how they can correct for multiple testing not just in statistical analyses but also in visualizations.

\begin{table}[h!]
    \centering
    \begin{tabular}{l|l|l|l}
         & $n = 40$ & $n = 120$ & $n = 200$  \\
         \hline
         Fail to reject chi-squared test at 0.05 level? & Yes & Yes & No \\
         \hline
         All 95\% CIs overlap? & Yes & Yes & No \\
         Fail to reject all pairwise tests at 0.05 level? & Yes & No & No \\
         \hline
         All 98.3\% CIs overlap? & Yes & Yes & Yes \\
         Fail to reject all pairwise tests at 0.05/3 level? & Yes & Yes & No \\
         \hline
    \end{tabular}
    \caption{Testing results for data displayed in Figure \ref{fig:barplots}.}
    \label{tab:testResults}
\end{table}

\re{It's worth noting that it's often critiqued to use bar plots with confidence intervals to display uncertainty \citep{newman2012bar,weissgerber2015beyond}; such visuals are sometimes referred to as ``dynamite plots'' or ``plunger plots.'' Dynamite plots use bars to display sample means and confidence intervals of a quantitative variable across different categories. Such plots are critiqued because, instead of displaying only sample means of a quantitative variable, it is more informative to visualize its distribution (e.g., via histograms, boxplots, or smoothed densities, as we note in the ``1D Quantitative \& 1D Categorical'' row in Table \ref{tab:taxonomy}). Meanwhile, Figure \ref{fig:barplots} displays just a single categorical variable: The height of the bars denotes the proportion in each category, rather than the sample mean of a quantitative variable, and thus would not be considered a dynamite plot.}

\re{Indeed, because bar plots visualize just a single categorical variable, many students may think that they are too basic for a college-level data visualization course. Figure \ref{fig:barplots} demonstrates that many inferential nuances nonetheless arise even in this simple setting.} This sets the stage for the rest of the semester, where students have to pinpoint which inferential tools are most appropriate for a particular graph. For example, when we teach visualizing quantitative distributions with overlayed densities, we acknowledge that there are many potential ways to compare distributions (e.g., inspecting means, variances, or distributional shapes), which correspond to many potential analyses (e.g., ANOVA, Bartlett's test, or the KS test). Once again, this introduces a multiple testing problem and issues of statistical power, because comparing shapes will have less power than comparing means. Students can more readily grasp these complications in more nuanced graphs like overlayed densities once they've seen a simpler example like Figure \ref{fig:barplots}. The primary takeaway is that, for almost any graph (simple or complex), students should consider how to determine if a graphical phenomenon is ``statistically significant,'' and what they can conclude from their graphs \textit{and} analyses together.

\re{Our discussion above focuses on inferential complications that can arise when examining a single graph. However, even more complications arise when we consider graphs and analyses made in sequence. For example, perhaps we conclude that population means of a quantitative variable differ across categories, based on a side-by-side boxplot and one-way ANOVA test; then, we may decide to create another graph and analysis to conduct another comparison. As a result, inference based on the second graph should be conditional on inference from the first graph. This is related to the literature on post-selection inference (e.g., \citealt{taylor2015statistical,tibshirani2016exact}), where one tries to conduct inference for a regression model conditional on certain variables being selected for that model (e.g., via stepwise selection)---but in this case, we must condition on results from previous graphs, instead of from selection procedures. Although this complication in exploratory data analyses has been acknowledged in various literatures \citep{buja2009statistical,zhao2017controlling,zgraggen2018investigating,savvides2022visual}, we do not discuss this in-depth in our class, in part because it is still an open problem how to validly conduct inference conditional on a sequence of visualizations. Nonetheless, grappling with this complication requires conceptualizing graphs as estimates and statistics that are used for inference---an idea we discuss next.}

\subsection{Principle 2: Graphs are estimates and statistics}

Figure \ref{fig:barplots} used a bar plot to display summary statistics (categorical proportions). Viewing these statistics as point estimates naturally motivated confidence intervals and hypothesis tests to complement the graph. More generally, graphs themselves can be viewed as estimates and statistics that can be leveraged within an inferential framework, such as uncertainty quantification and hypothesis testing.

First we discuss the pedagogical value of viewing graphs as estimates. For example, when we teach visualizing and analyzing two quantitative variables with scatterplots and linear regression, we emphasize that the regression line drawn on the scatterplot is indeed an estimator that has a distribution, which motivates displaying confidence bands for that line. One way to interpret 95\% confidence bands for a regression line is: If we obtained many random samples from the same population, and visualized each sample with a scatterplot and confidence bands, then approximately 95\% of those visual bands will contain the true line.\footnote{Technically, this interpretation is only true when one displays uniform confidence bands, rather than pointwise confidence bands. This inferential nuance is also a useful discussion point for students.} Although our undergraduate students are familiar with the definition of confidence intervals, this interpretation is often novel for them, \re{because it suggests a population-level graph and variation of a graph from sample to sample.}

\begin{figure}[h!]
    \centering
    \includegraphics[scale=0.7]{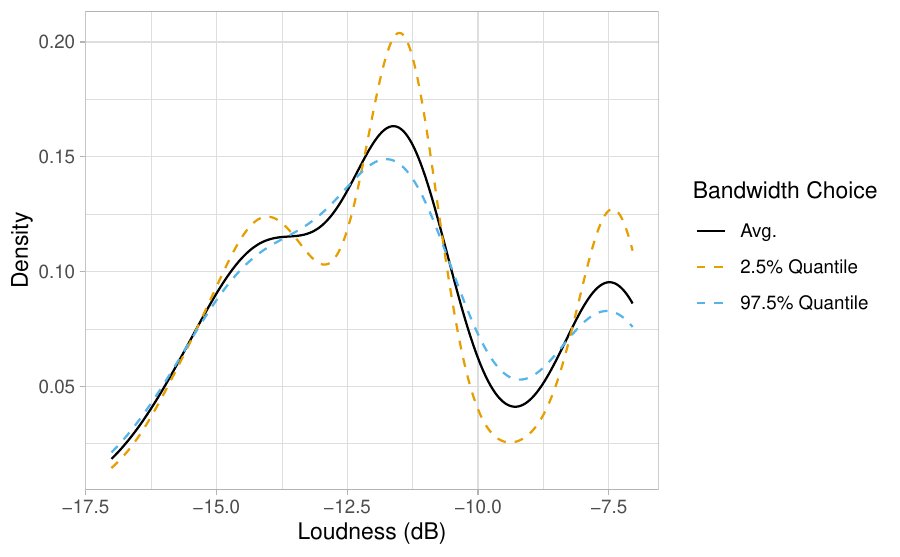}
    \caption{Distribution of loudness of songs, averaged within years from 1922-2021. Using this dataset of 100 observations, we generated 1000 bootstrapped datasets, and computed the estimated bandwidth for each bootstrapped dataset. Displayed is the density according to the average bandwidth (solid black), and the densities according to the lower 2.5\% quantile of bandwidths (dashed orange) and upper 97.5\% quantile bandwidth (dashed blue).}
    \label{fig:bandwidthPlot}
\end{figure}

To illustrate \re{how graphs can vary from sample to sample}, we'll consider visualizing a quantitative measure of loudness (in decibels) among songs on Spotify, averaged within years from 1922-2021.\footnote{The original dataset is available on Kaggle at \href{https://www.kaggle.com/datasets/yamaerenay/spotify-dataset-19212020-600k-tracks}{https://www.kaggle.com/datasets/yamaerenay/spotify-dataset-19212020-600k-tracks}; the dataset averaged across years is available in our replication materials.} Thus, the dataset consists of 100 observations, corresponding to the ``average'' song for a given year. Let's consider visualizing the distribution of loudness with a smoothed density, which involves estimating a bandwidth to quantify the distribution's smoothness. For example, the \texttt{density()} function in R uses Silverman's rule-of-thumb \citep{silverman2018density} by default for bandwidth estimation. This would lead to one smoothed density plot, but because the bandwidth is estimated, there is uncertainty in the distribution's smoothness and thus the plot itself. To understand this uncertainty, we created 1000 bootstrapped datasets via sampling with replacement, and computed the estimated bandwidth via Silverman's rule-of-thumb for each bootstrapped dataset. Then, we computed the average bandwidth, $\bar{h}$, across these 1000 bootstrapped datasets, as well as 2.5\% and 97.5\% quantiles, denoted $h_{(0.025)}$ and $h_{(0.975)}$. Figure \ref{fig:bandwidthPlot} displays three smoothed densities, corresponding to using $\bar{h}$, $h_{(0.025)}$, or $h_{(0.975)}$ as the bandwidth. The density using the average bandwidth $\bar{h}$ (in black) clearly has two modes on the right and middle of the graph, but it's less clear if there is a smaller third mode on the left. This smaller mode is more distinct when we use the smaller bandwidth $h_{(0.025)}$, but it's virtually nonexistent when we use the larger bandwidth $h_{(0.975)}$; meanwhile, the other two modes are clearly visible in both plots. This suggests that we have some uncertainty about whether this is a bimodal or trimodal distribution.

From a theoretical standpoint, this bootstrapping approach is not necessarily the optimal way to conduct inference for graphical properties like a distribution's smoothness and modes. For example, there is a rich literature on how to construct confidence sets for smoothed densities (e.g., \citealt[Chapter 6]{wasserman2006all}) and conduct inference for modes in a distribution (e.g., \citealt{genovese2016non}). \re{Furthermore, it's important to emphasize that this bootstrapping approach only illustrates variation in \textit{graphical properties}, like the bandwidth for smoothed densities, rather than the full variation in the graph itself. The reason is that graphical properties are often numeric summaries where averages and quantiles are well-understood, whereas it's more difficult to consider these operations on visuals. For example, Figure \ref{fig:bandwidthPlot} only visualizes three smoothed densities, resulting from the average bandwidth and its quantiles, and does not necessarily visualize the ``average'' graph or the ``quantiles'' of such graphs. Nonetheless, a bootstrap procedure is a useful way to demonstrate to students that graphs indeed have sample-to-sample variation.} For example, \cite{hesterberg2015teachers} discusses the pedagogical benefits of using a ``graphical bootstrap'' to illustrate the uncertainty of linear regression lines. Furthermore, bootstrapping procedures are relatively easy to implement and can flexibly incorporate many graphs, such that they have been used in recent statistical software to quantify uncertainty in data visualizations \citep{waskom2021seaborn,sun2023dynamic}. For students, the important takeaway is that graphs are statistics, because they are functions of observed data, just like sample means or medians. Graphs thus act as estimates for population-level quantities.

Because graphs are statistics, they can also be used as test statistics within a hypothesis testing procedure. In short, we can consider the ``null distribution'' of a graph under a given hypothesis, and view the graph we create for a given dataset as the observed test statistic. Using this insight, many works have proposed a ``lineup protocol,'' where the observed graph is plotted alongside graphs under a null distribution, and a test is conducted by determining if a user can differentiate the observed graph from graphs under the null \citep{buja2009statistical,wickham2010graphical,majumder2013validation,loy2016variations,vanderplas2020testing,loy2021bringing,li2024plot}. \cite{buja2009statistical} and \cite{vanderplas2020testing} discuss many examples of this procedure, such as testing temporal dependence by permuting and plotting many time series graphs. \cite{majumder2013validation} studied this procedure within the context of linear regression hypotheses (e.g., that a coefficient is equal to zero), and conducted several real experiments with human subjects to validate that this procedure has high statistical power. \re{\cite{loy2016variations} and \cite{li2024plot} found similar empirical evidence when using the lineup protocol for q-q plots and residual diagnostics, respectively. Meanwhile, \cite{loy2021bringing} discuss how lineup protocols are a useful pedagogical tool for teaching students the concepts of signal-versus-noise and variability.} As an illustrative example, we consider the lineup protocol for maps \citep{walter1993visual,wickham2010graphical}, to demonstrate how to apply this procedure to complex graphics, as we teach in the second half of our course.

\begin{figure}
\centering
     \includegraphics{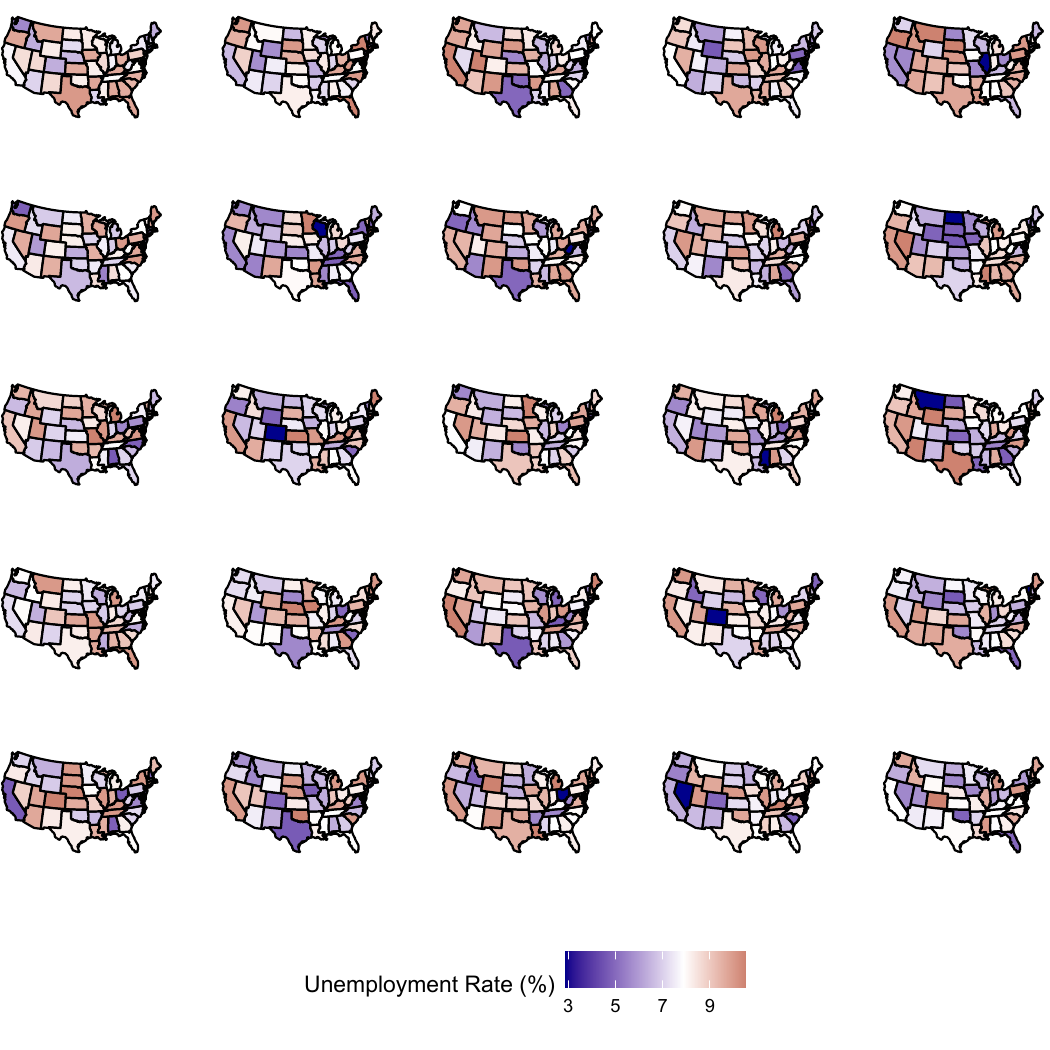}
    \caption{Average unemployment rate in each of the 48 contiguous United States, according to the 2015 American Community Survey. Displayed are 24 ``random maps'' where the data is randomly permuted across states, alongside the real map (which is in the second row, fifth column). The color coding is set such that white denotes the median unemployment rate across states.}
    \label{fig:maps}
\end{figure}

Figure \ref{fig:maps} illustrates how the lineup protocol operates for areal spatial data using choropleths, where geographic regions are colored to denote variable values. Using data from the 2015 American Community Survey, we plotted the average unemployment rate in each of the 48 contiguous United States.\footnote{The original dataset is available in our replication materials, as well as on Kaggle at \href{https://www.kaggle.com/datasets/census/2015-american-community-survey}{https://www.kaggle.com/datasets/census/2015-american-community-survey}. Variables are measured at the county level. To compute averages at the state level, we computed a weighted average, where weights corresponded to county population size.} Using the resulting map, we may ask: Is there spatial dependence in unemployment rates? We can answer this question from a testing perspective, where we consider what the distribution of this map is under the null hypothesis of spatial independence. To generate maps under this null hypothesis, we permuted the data across states and created a map with the resulting permuted data. Figure \ref{fig:maps} displays 24 of these maps alongside the real map; one conducts the hypothesis test by selecting the map that appears the most distinct among the 25 maps. Thus, if one selects the real map, the $p$-value is $1/25 = 0.04$, such that one would reject the null hypothesis of spatial independence at the $\alpha = 0.05$ level. Students are often surprised by the amount of apparent spatial clustering there can be in random maps; this gives them appreciation for the phenomenon that humans often find patterns in seemingly random visual stimuli, known as pareidolia.

Examples like Figures \ref{fig:bandwidthPlot} and \ref{fig:maps} illustrate to students that the concept of a ``distribution'' (whether it's across random samples or under a null hypothesis) can be broadly applied to any function of data. Others have found that bootstrapping and randomization test procedures can be a useful way to visually teach these concepts in introductory statistics courses \citep{cobb2007introductory,budgett2014students,hesterberg2015teachers}; these examples take this idea a step further by applying these procedures to visuals themselves. By teaching the principle that graphs are statistics, we've found that this helps students better understand the concept of a ``distribution,'' and provides students with tools to quantify the uncertainty in a given graph, especially complex ones like maps. As a result, students can more rigorously investigate whether a graphical finding is statistically significant, even when straightforward tests or models may not be readily available.

\subsection{Principle 3: Graphs can be used to motivate, teach, and interpret statistical analyses}

It's ubiquitous to use visuals to teach technical material, including statistical topics. For example, it's difficult to imagine teaching a course on linear regression without scatterplots, residual diagnostics, and other graphs. Thus, when teaching statistical analyses, it's well-known that graphs can help students gain a deeper understanding of those analyses. By the same logic, when teaching data visualization, instructors can use visuals to help students better understand statistical analyses they have seen and will see in future courses.

Indeed, one reason why we teach graphs and analyses in tandem is so that students can use graphs to quickly gain insights about technical material. For example, Table \ref{tab:taxonomy} illustrates how we use graphs like mosaic plots and side-by-side boxplots to discuss topics like chi-squared tests, $t$-tests, one-way ANOVA, and other analyses students may have seen before. In this sense, data visualization can act as a key part of spiral learning within an undergraduate curriculum, where students revisit topics several times at increasing levels of depth. To illustrate this idea, here we discuss two important statistical topics that are often challenging for students but are quite natural to learn within the context of data visualization: interactions in linear regression and principal component analysis.

To motivate interactions in regression, we'll consider a political science study assessing the impact of female and African American candidates on voter turnout \citep{keele2017black}. This study collected information on 1,006 mayoral elections in Louisiana from 1988-2011, including (1) whether at least one candidate in each election was female, (2) whether at least one candidate in each election was African American, (3) the unemployment rate of the municipality where each election was held, and (4) the voter turnout rate of each election. We may wonder whether voter turnout is associated with unemployment rates (which motivates regression), and whether this association is moderated by the presence of female and/or African American candidates (which motivates interactions within regression).

This study thus has two categorical variables and two quantitative variables. As is common in data visualization courses, we teach students how to display all four variables in a single scatterplot, where quantitative variables are displayed on the x- and y-axes and categorical variables are denoted by color and shape. Color and shape are known as ``aesthetics'' in the grammar of graphics framework, in the sense that data can be mapped to visual aesthetics \citep{wickham2010layered,wilkinson2012grammar}. The following example demonstrates that scatterplots can implicitly display different regression models, depending on how variables are aesthetically mapped to regression lines.

\begin{figure}[h!]
    \centering
    \includegraphics[scale=0.75]{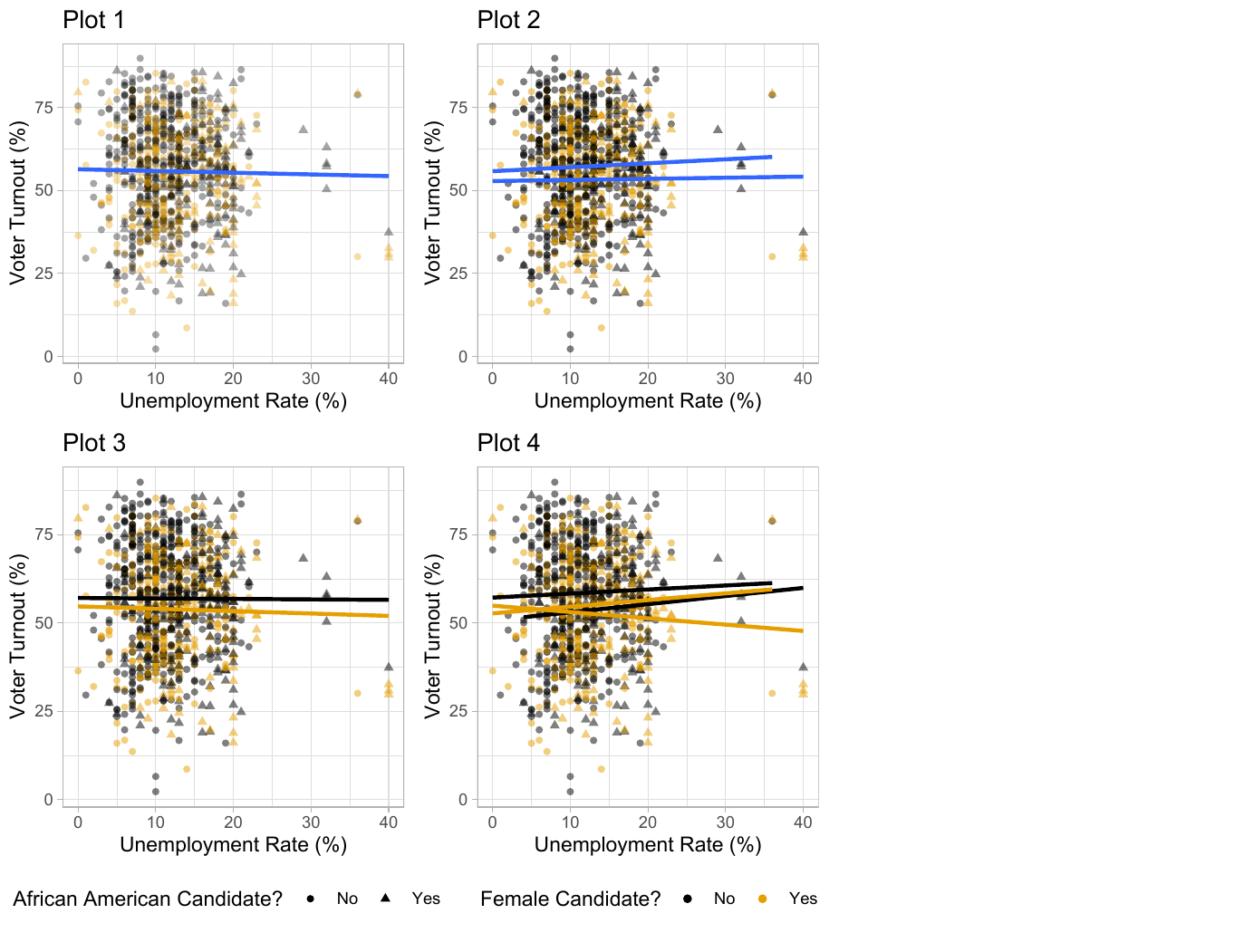}
    \caption{Scatterplots of municipal-level unemployment rates and voter turnout in Louisiana mayoral elections between 1988-2011. Points are colored and shaped by whether there was at least one female candidate and at least one African American candidate in a given election, respectively. The only difference among the plots is how points' color and shape are mapped to regression lines, which in turn corresponds to different statistical models. In Plot 1, no aesthetics are mapped to the regression line. In Plots 2 and 3, only the shape or color, respectively, are mapped to regression lines. In Plot 4, both shape and color are mapped to regression lines.}
    \label{fig:linInt}
\end{figure}

Figure \ref{fig:linInt} shows four scatterplots that display unemployment (x-axis) and voter turnout (y-axis), and use color and shape to denote the female and African American variables, respectively. The only difference among the plots is how color and shape are mapped to regression lines: No aesthetics are mapped to the regression in Plot 1 (resulting in one regression line), only color or shape are mapped in Plots 2 and 3 (resulting in two lines), and both color and shape are mapped in Plot 4 (resulting in four lines). Students often find these visual differences interesting because they correspond to subtle variations in code that many students tend to write. \re{However, this example demonstrates that once we map aesthetics to statistical models, inferential ambiguities can arise. For example, students can readily see that the plots in Figure \ref{fig:linInt} display regression models for different subsets of the data: Plot 1 shows a regression line for the entire dataset, and Plot 3 shows an orange regression line for elections with female candidates and a black regression line for elections without female candidates. Meanwhile, Plot 2 shows two regression lines---one for elections with African American candidates and one for those without---but it's ambiguous which line corresponds to which subset of data, because shapes cannot readily be mapped to lines. A similar issue arises with Plot 4. Given these plots, we ask students to resolve these ambiguities by having them implement the regression models implicitly displayed by each plot.} For example, Table \ref{tab:linInt} shows the estimated coefficients for a model that regresses voter turnout on unemployment, the female and African American variables, and their interactions, which corresponds to the regression displayed in Plot 4. \re{Students can then use this output to determine which regression line corresponds to which subset of data by interpreting the estimated coefficients in terms of changes in intercepts and slopes. For example, because the estimated coefficient for the three-way interaction is negative (and is greater in magnitude than the estimated coefficients for the two-way interactions involving unemployment), we can infer that the downward-sloping orange line corresponds to the regression for elections that have African American and female candidates. We've found that students can more readily realize that interactions in linear regression models correspond to changes in intercepts and slopes when they interpret output like Table \ref{tab:linInt} alongside its corresponding graph.}

\re{Grappling with the inferential ambiguities in Figure \ref{fig:linInt} can also help students appreciate the implications---and potential downsides---of certain visual design choices. Humans tend to focus on differences in color over differences in shape \citep{wagemans2012century,garner2014processing}, such that Figure \ref{fig:linInt} implicitly emphasizes the orange-black differences over the circle-triangle differences. By pinpointing which statistical model in Figure \ref{fig:linInt} we'd like to emphasize, we can determine how to improve a plot's graphic design. For example, if we wanted to utilize a model that only considers differences by the African American variable (as in Plot 2) or by the female variable (as in Plot 3), we may only use color for the variable of interest and keep all observations the same shape. On the other hand, if we wanted to utilize a model that considers the interaction between these variables (as in Plot 4), we may use four colors to denote these four groups, or we may facet by one categorical variable to create two scatterplots that each use color to differentiate observations.}

There are three intended lessons from this example. First, there is often an implicit statistical model suggested by a graph, depending on how variables are mapped to the graph's aesthetics. It can be useful to have students consider which model is suggested by a particular graph, and then implement that model; this lets the graph motivate the analysis, which is often intuitive for students. Second, by placing graphs and their corresponding analyses side-by-side, students can more quickly understand and interpret statistical analyses by seeing how they operate graphically. \re{Third, by assessing whether a graphic aligns with an intended statistical model, students can pinpoint how to improve their graphs' visual design.} This brings students closer to being effective data scientists, who use graphs and analyses synergistically to arrive at comprehensive conclusions.

\begin{table}[h!]
    \centering
    \begin{tabular}{|c|c|}
    \hline
       \textbf{Variable(s)}  & \textbf{Estimated Coefficients}  \\
       \hline
        (Intercept) & 57.15 \\
        Unemployment & 0.11 \\
        Female (yes) & -4.46 \\
        African American (yes) & -6.47 \\
        Female (yes):African American (yes) & 8.64 \\
        Unemployment:female (yes) & 0.07 \\
        Unemployment:African American (yes) & 0.12 \\
        Unemployment:female (yes):African American (yes) & -0.48 \\
        \hline
    \end{tabular}
    \caption{Estimated regression coefficients from the linear regression model implied by Plot 4 in Figure \ref{fig:linInt}, which regresses voter turnout on unemployment, the female and African American variables, and their interactions (denoted by colons).}
    \label{tab:linInt}
\end{table}

Finally, as an example that motivates principal component analysis (PCA), we'll revisit the Spotify dataset from the previous subsection, where we visualized the loudness of songs in Figure \ref{fig:bandwidthPlot}. In addition to loudness, there are 10 measures: a song's ``acousticness,'' ``danceability,'' length, energy, ``instrumentalness,'' ``liveness,'' popularity, ``speechiness,'' tempo, and valence.\footnote{Details about how these are measured are described in the aforementioned Kaggle link.} Thus, this dataset lends itself well to high-dimensional data visualizations, because it's seemingly challenging to visualize 11 dimensions in a single plot.

PCA is a common dimension-reduction technique for visualization and statistical analyses. There are two PCA-based visuals that many students are familiar with. First, a scree plot (or ``elbow plot'') can be used to choose the number of principal components. Second, a scatterplot of principal components can be used to identify underlying clusters in the data. However, we've found that many students have trouble interpreting the chosen number of components in terms of visualizations (e.g., ``I think I should use four components... so what visuals do I make?''), and have trouble interpreting the components themselves or resulting clusters (e.g., ``I see clusters in my scatterplot, but I don't know how to describe these clusters''). To help students address these questions, we've found it useful to examine quantitative output from PCA alongside visualizations and ask students how the output and visualizations can complement each other.

\begin{figure}[h!]
    \centering
    \begin{subfigure}[t]{0.45\textwidth}
    \includegraphics[scale=0.5]{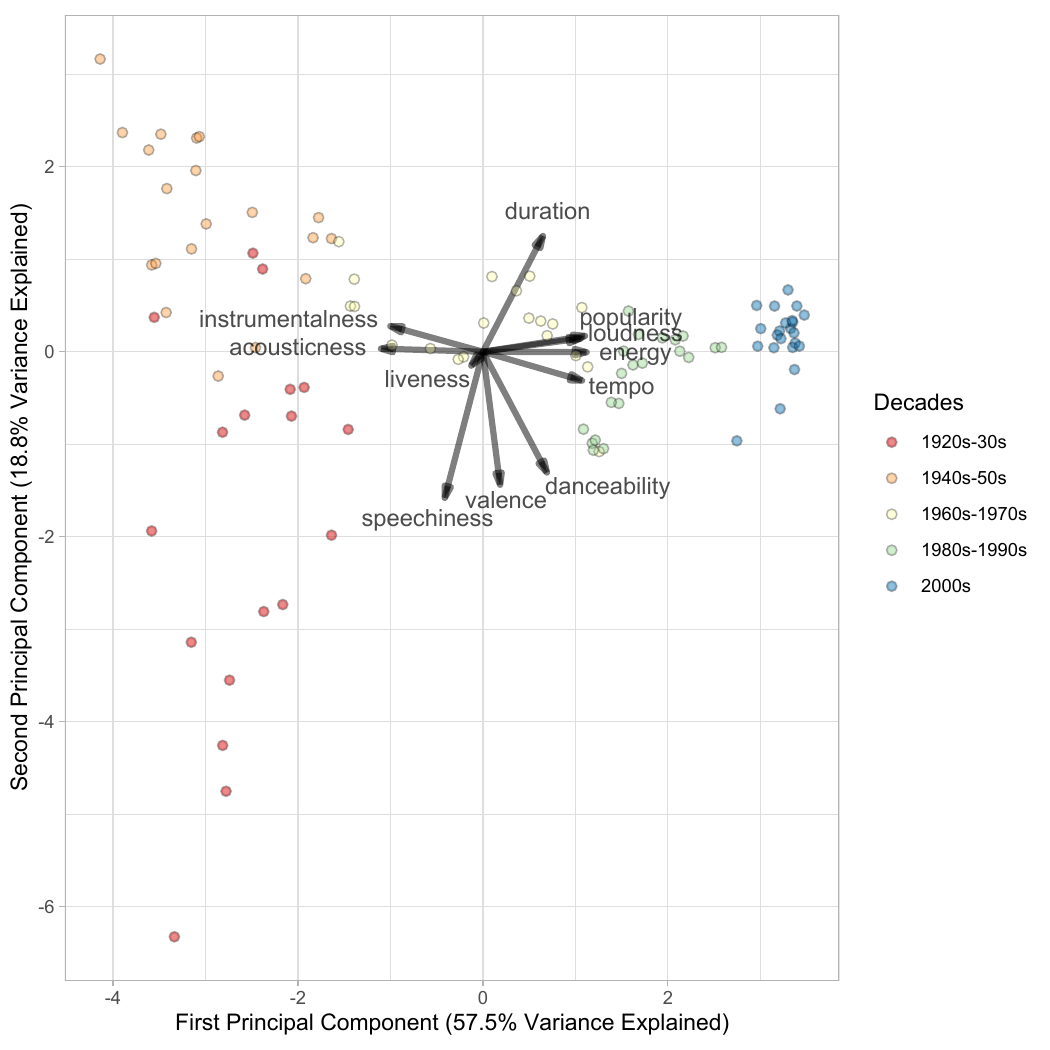}
    \caption{Biplot of the top two principal components and 11 variables. Color denotes decades that songs were released. Length and direction of arrows reflect variables' relationship with the principal components.}
    \label{fig:biplot}
    \end{subfigure}
    \hfill
    \begin{subfigure}[t]{0.45\textwidth}
        \includegraphics[scale=0.5]{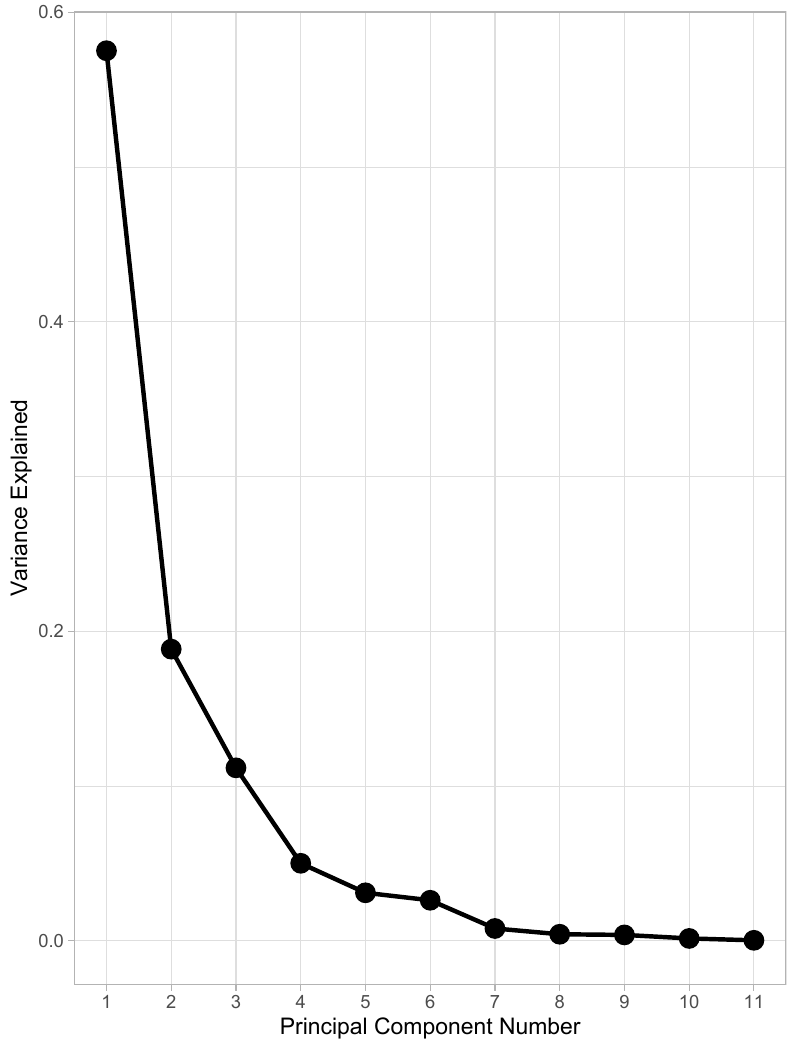}
        \caption{Scree plot. Variance explained (y-axis) is defined as each principal component's variance divided by the sum of variances. One can argue that the ``elbow'' is anywhere between four and seven principal components.}
        \label{fig:elbowPlot}
    \end{subfigure}
    \caption{Biplot (left) and scree plot (right) for the Spotify data.}
\end{figure}

Because our course is meant for a broad set of students, we avoid using matrix algebra to explain PCA. Instead, we emphasize two facts: (1) the components are defined as linear combinations of the (scaled) variables (known as ``loadings''), and (2) the first component has the highest variance, followed by the second, and so on, such that the sum of variances is equal to the number of variables. Because we cover linear regression earlier in the course, students are familiar with linear combinations of variables. For example, Table \ref{tab:pca} shows the loadings for the first two components for the Spotify data. When we display the components on a scatterplot, the loadings help students understand where the dots in the plot come from: we multiply each variable's value by its loading coefficient and add, just like we do to obtain predicted values in linear regression. Figure \ref{fig:biplot} shows such a scatterplot, with the first principal component on the x-axis, the second principal component on the y-axis, and dots colored by the decades that songs were released (the only categorical variable in the dataset). If we ignore the arrows in Figure \ref{fig:biplot} for a moment (which we will discuss shortly), we can see that, interestingly, songs become more recent as we go from left to right in the plot. At first, students are often unsure how to interpret this phenomenon, because they're unsure how to interpret the principal components. We then ask students to revisit the loadings in Table \ref{tab:pca}, and consider how these quantitative results can be incorporated into the plot.

A biplot \citep{gabriel1971biplot} is particularly useful for this task, which in its simplest form overlays the scatterplot with arrows visualizing these loadings, as displayed in Figure \ref{fig:biplot}. There is an arrow for each variable; the length and direction of the arrows denote the magnitude and sign, respectively, of the loadings, thereby visualizing the variables' relationship with these two principal components. \re{This allows us to interpret observations in the principal component space in terms of the original variables in the data. For example, in Figure \ref{fig:biplot}, observations on the left tend to have higher levels of ``instrumentalness'' and ``acousticness,'' whereas observations on the right tend to have higher levels of popularity, loudness, energy, and tempo. Because songs become more recent as we go from left to right, this suggests that more recent songs tend to have a higher tempo, energy, loudness, and popularity, while older songs tend to have higher levels of ``instrumentalness'' and ``acousticness.'' That said, there may be other unmeasured variabes (besides decade of release) that differentiate observations from left-to-right in the plot. For example, we could have taken an unsupervised approach by applying $k$-means clustering to the observations in this two-dimensional space, and then used the loadings to interpret the resulting clusters, without reference to decade of release.} In any case, the biplot can help students gain deeper insights about visualizations (e.g., how to interpret scatterplots of principal components) as well as analyses (e.g., understanding what the loadings communicate substantively).

\re{The biplot in Figure \ref{fig:biplot} only partially captures potential clustering of observations, because it's projecting a 10-dimensional space down to two dimensions. This is where a scree plot can be helpful for understanding how much variation in the data we're ignoring when interpreting the biplot.} Figure \ref{fig:elbowPlot} shows a scree plot for the Spotify data. It's somewhat subjective where the ``elbow'' in the plot is; one can argue that it is anywhere between $k = 4$ and $k = 7$ components. Instead of only focusing on the ``elbow,'' we ask students to consider what the ``variance explained'' (y-axis) means in terms of scatterplots of components. Within a scatterplot, a component's variance corresponds to horizontal or vertical spread. Thus, plotting the first two components corresponds to the scatterplot with the most horizontal and vertical spread---in this case, $57.5\% + 18.8\% = 76.3\%$ of the total spread. If we decided to use additional components, we would make additional scatterplots, but those scatterplots would have much less visual spread than the one in Figure \ref{fig:biplot}. \re{That said, Figure \ref{fig:biplot} ignores 23.7\% of the variation in the data, which could reveal other potential clusters if we examined it with further visualizations.} This helps students understand the concept of ``variance explained'' within PCA in terms of visual spread, and why organizing components by variance is useful for visualization.

\begin{table}[h!]
    \centering
    \begin{tabular}{c|c|c}
\hline
  \textbf{Variable} & \textbf{First Principal Component} & \textbf{Second Principal Component} \\
\hline
acousticness & -0.388 & 0.012\\
\hline
danceability & 0.243 & -0.459\\
\hline
duration & 0.228 & 0.440\\
\hline
energy & 0.394 & -0.002\\
\hline
instrumentalness & -0.352 & 0.097\\
\hline
liveness & -0.045 & -0.053\\
\hline
loudness & 0.372 & 0.050\\
\hline
popularity & 0.388 & 0.060\\
\hline
speechiness & -0.145 & -0.556\\
\hline
tempo & 0.377 & -0.110\\
\hline
valence & 0.065 & -0.506\\
\hline
\end{tabular}
    \caption{Loadings for the first two principal components in the Spotify data.}
    \label{tab:pca}
\end{table}

PCA, biplots, and interactions in regression are all classical statistical techniques, and certainly it is not novel that we teach these topics in a statistics course. Rather, our intended lesson here is that data visualization courses present a unique opportunity to discuss fairly technical material in a way that is accessible for students, by using graphs as a facilitative tool. In this way, data visualization courses can act as a useful conduit for teaching statistical thinking. This also encourages spiral learning within a curriculum, where ideally students utilize visual insights when they engage with technical material later in their careers.

\section{Conclusion and Discussion} \label{s:conclusion}

Data visualization has a long history in statistics and is ubiquitous across many fields and professions, from the sciences to the arts. Although data visualization is frequently used in many statistics courses, it's usually not the focus of the course---more often, it's used as a tool to teach other concepts. Thus, instructors and departments may wonder if it's beneficial to teach data visualization itself. Indeed, visualizing data is arguably a core skill of any statistical practice. That said, because data visualization is such a broad discipline, some may be unsure how to teach it or what topics to focus on. To give guidance on teaching data visualization from a statistical perspective, we've made two contributions.

First, we conducted a survey of data visualization courses taught at 135 top colleges and universities in the United States in the 2022-2023 academic year, in order to understand the landscape of data visualization courses. We identified 270 courses across many departments; the most common were related to computer science, data, information, and management, but departments in analytics, business, engineering, the humanities, journalism, public policy, statistics, the visual arts, and other fields were also represented. Thus, the instruction of data visualization is widespread and diverse. Most courses were not taught by statistics or data science departments, and thus understandably did not focus on statistical topics. Instead, most courses focused on visual storytelling, aesthetic design, dashboard design, specialized software, or other topics unique to their respective disciplines. It was especially uncommon for courses to focus on core aspects of statistical thinking: testing, uncertainty quantification, and modeling. Thus, statistics-related departments can provide a valuable perspective on data visualization that is unique to current course offerings.

Second, we outlined three principles for teaching data visualization to encourage statistical thinking, using illustrative examples from our undergraduate data visualization course at Carnegie Mellon University. The first principle is that graphs should be complemented with inference. For example, the ``statistical significance'' of visual observations (e.g., differences in proportions or means) can be clarified with standard tests (e.g., chi-squared tests or ANOVA). This helps students understand how to combine graphs and analyses to make a cohesive argument, and encourages them to grapple with complex inferential issues (e.g., power, multiple testing, and visually overlapping confidence intervals). The second principle is that graphs are estimates and statistics that can be leveraged in an inferential framework. For example, graphs can have a null distribution that's used in a testing procedure, and have uncertainty that can be quantified theoretically or via simulation with a bootstrap procedure. This helps students gain a deeper understanding of distributions and estimators by seeing how they operate with complex visuals beyond just univariate random variables. The third principle is that graphs can be used to motivate, teach, and interpret statistical analyses. Because visuals are a tried-and-true teaching tool, data visualization courses can be a natural conduit for students to engage with challenging concepts, like interactions in regression and principal component analysis. In this way, we've used data visualization as part of a spiral learning approach in our undergraduate curriculum, where students revisit topics several times (e.g., in a visualization context and in a theoretical context) as they master difficult material.

That said, there are several limitations to our contributions, which motivate future work. Our survey is likely not a census of all data visualization courses, because (1) we used select keywords to identify potential courses, (2) we relied only on publicly available information to declare a course as a data visualization course, and (3) we only considered top colleges and universities according to the U.S. News and World Report in 2023. \re{Furthermore, because our survey only considers courses taught in the 2022-2023 academic year, we have not considered how instruction on data visualization has possibly changed over time.} Finally, we focused primarily on courses' statistical content and did not form a full taxonomy of topics taught in data visualization courses. For example, we observed that many courses focused on visual storytelling and aesthetic design (particularly in business, journalism, the visual arts, and writing), but did not study which methods within storytelling and design these courses taught. Because different disciplines have their own specialized interest in data visualization, other fields may want to study which topics are covered in data visualization courses within their own discipline, perhaps using our publicly available dataset on these courses. Indeed, we hope our dataset allows others to explore the diversity of data visualization courses, including instructors and students seeking datasets for undergraduate projects.

Furthermore, while we hope our three principles give guidance on how to teach data visualization from a statistical perspective, we do not claim that effective courses need to follow these principles, even within statistics. For example, statistics and data science departments may rightfully want to highlight visual storytelling, aesthetic design, dashboard design, specialized software, or other topics. One observation from our survey is that most data visualization courses focus on these topics, whereas very few touch on statistical inference. Thus, our principles provide a way for instructors to incorporate inferential ideas into data visualization courses. In the supplementary material we give details about how we design our courses at the undergraduate and graduate level, so that instructors and departments can see how to design courses with these principles in mind.

Finally, there are likely many other ways to incorporate statistical thinking and inference in data visualization courses that we do not discuss here. We hope that this work encourages departments in statistics, data science, and related fields to consider how data visualization courses can encourage statistical thinking in their own curricula, especially because statistical thinking is not widely emphasized in current courses. In this way, statistics and data science can take more ownership of modern data visualization, while making meaningful contributions to college-level courses.

\if0\blind
{
\section*{Acknowledgments}

We would like to thank Peter Freeman, Alex Reinhart, and three anonymous reviewers for helpful comments that improved the presentation of this work.
}\fi

\bibliographystyle{apalike}
\bibliography{statsGraphicsBib}

\begin{thebibliography}{}

\bibitem[Anscombe, 1973]{anscombe1973graphs}
Anscombe, F.~J. (1973).
\newblock Graphs in statistical analysis.
\newblock {\em The {A}merican {S}tatistician}, 27(1):17--21.

\bibitem[Bruner, 2009]{bruner2009process}
Bruner, J.~S. (2009).
\newblock {\em The {P}rocess of {E}ducation}.
\newblock Harvard {U}niversity {P}ress.

\bibitem[Budgett and Wild, 2014]{budgett2014students}
Budgett, S. and Wild, C.~J. (2014).
\newblock Students' visual reasoning and the randomization test.
\newblock In {\em Proceedings of the 9th International Conference on Teaching
  Statistics}.

\bibitem[Buja et~al., 2009]{buja2009statistical}
Buja, A., Cook, D., Hofmann, H., Lawrence, M., Lee, E.-K., Swayne, D.~F., and
  Wickham, H. (2009).
\newblock Statistical inference for exploratory data analysis and model
  diagnostics.
\newblock {\em Philosophical Transactions of the Royal Society A: Mathematical,
  Physical and Engineering Sciences}, 367(1906):4361--4383.

\bibitem[Cairo, 2012]{cairo2012functional}
Cairo, A. (2012).
\newblock {\em The Functional Art: An {I}ntroduction to {I}nformation
  {G}raphics and {V}isualization}.
\newblock New Riders.

\bibitem[Cairo, 2016]{cairo2016truthful}
Cairo, A. (2016).
\newblock {\em The {T}ruthful {A}rt: Data, {C}harts, and {M}aps for
  {C}ommunication}.
\newblock New Riders.

\bibitem[Cairo, 2019]{cairo2019charts}
Cairo, A. (2019).
\newblock {\em How {C}harts {L}ie: Getting {S}marter about {V}isual
  {I}nformation}.
\newblock WW Norton \& Company.

\bibitem[Cairo, 2023]{cairo2023art}
Cairo, A. (2023).
\newblock {\em The Art of Insight: How Great Visualization Designers Think}.
\newblock John Wiley \& Sons.

\bibitem[Carver et~al., 2016]{carver2016guidelines}
Carver, R., Everson, M., Gabrosek, J., Horton, N., Lock, R., Mocko, M.,
  Rossman, A., Roswell, G.~H., Velleman, P., and Witmer, J. (2016).
\newblock Guidelines for {A}ssessment and {I}nstruction in {S}tatistics
  {E}ducation ({GAISE}) {C}ollege {R}eport 2016.

\bibitem[Chatterjee and Firat, 2007]{chatterjee2007generating}
Chatterjee, S. and Firat, A. (2007).
\newblock Generating data with identical statistics but dissimilar graphics: A
  follow up to the anscombe dataset.
\newblock {\em The American Statistician}, 61(3):248--254.

\bibitem[Chen et~al., 2007]{chen2007handbook}
Chen, C.-h., H{\"a}rdle, W.~K., and Unwin, A. (2007).
\newblock {\em Handbook of {D}ata {V}isualization}.
\newblock Springer Science \& Business Media.

\bibitem[Chernoff, 1973]{chernoff1973use}
Chernoff, H. (1973).
\newblock The use of faces to represent points in k-dimensional space
  graphically.
\newblock {\em Journal of the American Statistical Association},
  68(342):361--368.

\bibitem[Cobb, 2007]{cobb2007introductory}
Cobb, G.~W. (2007).
\newblock The introductory statistics course: A ptolemaic curriculum?
\newblock {\em Technology Innovations in Statistics Education}, 1(1).

\bibitem[Cohen and Cohen, 2006]{cohen2006sectioned}
Cohen, D.~J. and Cohen, J. (2006).
\newblock The sectioned density plot.
\newblock {\em The American Statistician}, 60(2):167--174.

\bibitem[Cumming and Finch, 2005]{cumming2005inference}
Cumming, G. and Finch, S. (2005).
\newblock Inference by eye: confidence intervals and how to read pictures of
  data.
\newblock {\em American Psychologist}, 60(2):170.

\bibitem[Dogucu and Hu, 2022]{dogucu2022current}
Dogucu, M. and Hu, J. (2022).
\newblock The current state of undergraduate bayesian education and
  recommendations for the future.
\newblock {\em The American Statistician}, 76(4):405--413.

\bibitem[Engebretsen and Kennedy, 2020]{engebretsen2020data}
Engebretsen, M. and Kennedy, H. (2020).
\newblock {\em Data {V}isualization in {S}ociety}.
\newblock Amsterdam {U}niversity {P}ress.

\bibitem[Few, 2004]{few2004show}
Few, S. (2004).
\newblock {\em Show Me the Numbers}, volume~2.
\newblock Analytics Press.

\bibitem[Friendly, 2008]{friendly2008}
Friendly, M. (2008).
\newblock A brief history of data visualization.
\newblock In {\em Handbook of Data Visualization}, pages 15--56. Springer.

\bibitem[Gabriel, 1971]{gabriel1971biplot}
Gabriel, K.~R. (1971).
\newblock The biplot graphic display of matrices with application to principal
  component analysis.
\newblock {\em Biometrika}, 58(3):453--467.

\bibitem[Garner, 2014]{garner2014processing}
Garner, W.~R. (2014).
\newblock {\em The Processing of Information and Structure}.
\newblock Psychology Press.

\bibitem[Gelman and Unwin, 2013]{gelman2013infovis}
Gelman, A. and Unwin, A. (2013).
\newblock Infovis and statistical graphics: different goals, different looks.
\newblock {\em Journal of Computational and Graphical Statistics}, 22(1):2--28.

\bibitem[Genovese et~al., 2016]{genovese2016non}
Genovese, C.~R., Perone-Pacifico, M., Verdinelli, I., and Wasserman, L. (2016).
\newblock Non-parametric inference for density modes.
\newblock {\em Journal of the Royal Statistical Society Series B: Statistical
  Methodology}, 78(1):99--126.

\bibitem[Harden, 1999]{harden1999spiral}
Harden, R.~M. (1999).
\newblock What is a spiral curriculum?
\newblock {\em Medical Teacher}, 21(2):141--143.

\bibitem[Healy, 2018]{healy2018data}
Healy, K. (2018).
\newblock {\em Data {V}isualization: {A} {P}ractical {I}ntroduction}.
\newblock Princeton University Press.

\bibitem[Hesterberg, 2015]{hesterberg2015teachers}
Hesterberg, T.~C. (2015).
\newblock What teachers should know about the bootstrap: Resampling in the
  undergraduate statistics curriculum.
\newblock {\em The American Statistician}, 69(4):371--386.

\bibitem[Hintze and Nelson, 1998]{hintze1998violin}
Hintze, J.~L. and Nelson, R.~D. (1998).
\newblock Violin plots: a box plot-density trace synergism.
\newblock {\em The American Statistician}, 52(2):181--184.

\bibitem[Jones, 2014]{jones2014communicating}
Jones, B. (2014).
\newblock {\em Communicating {D}ata with {T}ableau: Designing, {D}eveloping,
  and {D}elivering {D}ata {V}isualizations}.
\newblock O'Reilly Media, Inc.

\bibitem[Keele et~al., 2017]{keele2017black}
Keele, L.~J., Shah, P.~R., White, I., and Kay, K. (2017).
\newblock Black candidates and black turnout: A study of viability in louisiana
  mayoral elections.
\newblock {\em The Journal of Politics}, 79(3):780--791.

\bibitem[Kleiner and Hartigan, 1981]{kleiner1981representing}
Kleiner, B. and Hartigan, J.~A. (1981).
\newblock Representing points in many dimensions by trees and castles.
\newblock {\em Journal of the American Statistical Association},
  76(374):260--269.

\bibitem[Knaflic, 2015]{knaflic2015storytelling}
Knaflic, C.~N. (2015).
\newblock {\em Storytelling with {D}ata: A {D}ata {V}isualization {G}uide for
  {B}usiness {P}rofessionals}.
\newblock John Wiley \& Sons.

\bibitem[Li et~al., 2024]{li2024plot}
Li, W., Cook, D., Tanaka, E., and VanderPlas, S. (2024).
\newblock A plot is worth a thousand tests: Assessing residual diagnostics with
  the lineup protocol.
\newblock {\em Journal of Computational and Graphical Statistics},
  33(4):1497--1511.

\bibitem[Loy, 2021]{loy2021bringing}
Loy, A. (2021).
\newblock Bringing visual inference to the classroom.
\newblock {\em Journal of Statistics and Data Science Education},
  29(2):171--182.

\bibitem[Loy et~al., 2016]{loy2016variations}
Loy, A., Follett, L., and Hofmann, H. (2016).
\newblock Variations of q--q plots: The power of our eyes!
\newblock {\em The American Statistician}, 70(2):202--214.

\bibitem[Majumder et~al., 2013]{majumder2013validation}
Majumder, M., Hofmann, H., and Cook, D. (2013).
\newblock Validation of visual statistical inference, applied to linear models.
\newblock {\em Journal of the American Statistical Association},
  108(503):942--956.

\bibitem[Matejka and Fitzmaurice, 2017]{matejka2017same}
Matejka, J. and Fitzmaurice, G. (2017).
\newblock Same stats, different graphs: generating datasets with varied
  appearance and identical statistics through simulated annealing.
\newblock In {\em Proceedings of the 2017 CHI Conference on Human Factors in
  Computing Systems}, pages 1290--1294.

\bibitem[Murray, 2017]{murray2017interactive}
Murray, S. (2017).
\newblock {\em Interactive Data Visualization for the Web: An Introduction to
  Designing with D3}.
\newblock O'Reilly Media, Inc.

\bibitem[Newman and Scholl, 2012]{newman2012bar}
Newman, G.~E. and Scholl, B.~J. (2012).
\newblock Bar graphs depicting averages are perceptually misinterpreted: The
  within-the-bar bias.
\newblock {\em Psychonomic Bulletin \& Review}, 19:601--607.

\bibitem[Nolan and Stoudt, 2021]{nolan2021promise}
Nolan, D. and Stoudt, S. (2021).
\newblock The promise of portfolios: Training modern data scientists.
\newblock {\em Harvard Data Science Review}, 3(3).

\bibitem[Reiter, 2023]{reiter2023data}
Reiter, A.~G. (2023).
\newblock {U.S.} news \& world report historical liberal arts college and
  university rankings.
\newblock Accessed June 1, 2023, URL http://andyreiter.com/datasets/.

\bibitem[Sarkar, 2008]{sarkar2008multivariate}
Sarkar, D. (2008).
\newblock Multivariate data visualization with {R}.
\newblock {\em Use R}.

\bibitem[Savvides et~al., 2022]{savvides2022visual}
Savvides, R., Henelius, A., Oikarinen, E., and Puolam{\"a}ki, K. (2022).
\newblock Visual data exploration as a statistical testing procedure:
  Within-view and between-view multiple comparisons.
\newblock {\em IEEE Transactions on Visualization and Computer Graphics},
  29(9):3937--3948.

\bibitem[Schwabish, 2021]{schwabish2021better}
Schwabish, J. (2021).
\newblock {\em Better Data Visualizations: A Guide for Scholars, Researchers,
  and Wonks}.
\newblock Columbia University Press.

\bibitem[Sievert, 2020]{sievert2020interactive}
Sievert, C. (2020).
\newblock {\em Interactive Web-based Data Visualization with R, Plotly, and
  Shiny}.
\newblock CRC Press.

\bibitem[Silverman, 2018]{silverman2018density}
Silverman, B.~W. (2018).
\newblock {\em Density Estimation for Statistics and Data Analysis}.
\newblock Routledge.

\bibitem[Soukup and Davidson, 2002]{soukup2002visual}
Soukup, T. and Davidson, I. (2002).
\newblock {\em Visual Data Mining: Techniques and Tools for Data Visualization
  and Mining}.
\newblock John Wiley \& Sons.

\bibitem[Sun et~al., 2023]{sun2023dynamic}
Sun, E.~D., Ma, R., and Zou, J. (2023).
\newblock Dynamic visualization of high-dimensional data.
\newblock {\em Nature Computational Science}, 3(1):86--100.

\bibitem[Taylor and Tibshirani, 2015]{taylor2015statistical}
Taylor, J. and Tibshirani, R.~J. (2015).
\newblock Statistical learning and selective inference.
\newblock {\em Proceedings of the National Academy of Sciences},
  112(25):7629--7634.

\bibitem[Tibshirani et~al., 2016]{tibshirani2016exact}
Tibshirani, R.~J., Taylor, J., Lockhart, R., and Tibshirani, R. (2016).
\newblock Exact post-selection inference for sequential regression procedures.
\newblock {\em Journal of the American Statistical Association},
  111(514):600--620.

\bibitem[Tufte, 2001]{tufte2001visual}
Tufte, E.~R. (2001).
\newblock {\em The Visual Display of Quantitative Information}, volume~2.
\newblock Graphics press Cheshire, CT.

\bibitem[Tukey, 1977]{tukey1977exploratory}
Tukey, J.~W. (1977).
\newblock {\em Exploratory Data Analysis}.
\newblock Reading/Addison-Wesley.

\bibitem[Unwin, 2024]{unwin2024getting}
Unwin, A. (2024).
\newblock {\em Getting (More Out Of) Graphics: Practice and Principles of Data
  Visualisation}.
\newblock Chapman and Hall/CRC.

\bibitem[{U.S. News}, 2023a]{us2023universities}
{U.S. News} (2023a).
\newblock Best national university rankings.
\newblock Accessed June 1, 2023, URL
  https://www.usnews.com/best-colleges/rankings/national-universities.

\bibitem[{U.S. News}, 2023b]{us2023colleges}
{U.S. News} (2023b).
\newblock National liberal arts colleges.
\newblock Accessed June 1, 2023, URL
  https://www.usnews.com/best-colleges/rankings/national-liberal-arts-colleges.

\bibitem[Vanderplas et~al., 2020]{vanderplas2020testing}
Vanderplas, S., Cook, D., and Hofmann, H. (2020).
\newblock Testing statistical charts: What makes a good graph?
\newblock {\em Annual Review of Statistics and Its Application}, 7(1):61--88.

\bibitem[Wagemans et~al., 2012]{wagemans2012century}
Wagemans, J., Feldman, J., Gepshtein, S., Kimchi, R., Pomerantz, J.~R., Van~der
  Helm, P.~A., and Van~Leeuwen, C. (2012).
\newblock A century of gestalt psychology in visual perception: Ii. conceptual
  and theoretical foundations.
\newblock {\em Psychological Bulletin}, 138(6):1218.

\bibitem[Walter, 1993]{walter1993visual}
Walter, S. (1993).
\newblock Visual and statistical assessment of spatial clustering in mapped
  data.
\newblock {\em Statistics in {M}edicine}, 12(14):1275--1291.

\bibitem[Waskom, 2021]{waskom2021seaborn}
Waskom, M.~L. (2021).
\newblock Seaborn: {S}tatistical data visualization.
\newblock {\em Journal of Open Source Software}, 6(60):3021.

\bibitem[Wasserman, 2006]{wasserman2006all}
Wasserman, L. (2006).
\newblock {\em All of Nonparametric Statistics}.
\newblock Springer Science \& Business Media.

\bibitem[Weissgerber et~al., 2015]{weissgerber2015beyond}
Weissgerber, T.~L., Milic, N.~M., Winham, S.~J., and Garovic, V.~D. (2015).
\newblock Beyond bar and line graphs: time for a new data presentation
  paradigm.
\newblock {\em PLoS Biology}, 13(4):e1002128.

\bibitem[Wickham, 2010]{wickham2010layered}
Wickham, H. (2010).
\newblock A layered grammar of graphics.
\newblock {\em Journal of Computational and Graphical Statistics}, 19(1):3--28.

\bibitem[Wickham, 2016]{wickham2016data}
Wickham, H. (2016).
\newblock {\em ggplot2: Elegant Graphics for Data Analysis}.
\newblock Springer.

\bibitem[Wickham et~al., 2019]{wickham2019welcome}
Wickham, H., Averick, M., Bryan, J., Chang, W., McGowan, L.~D., Fran{\c{c}}ois,
  R., Grolemund, G., Hayes, A., Henry, L., Hester, J., et~al. (2019).
\newblock Welcome to the tidyverse.
\newblock {\em Journal of Open Source Software}, 4(43):1686.

\bibitem[Wickham et~al., 2010]{wickham2010graphical}
Wickham, H., Cook, D., Hofmann, H., and Buja, A. (2010).
\newblock Graphical inference for infovis.
\newblock {\em IEEE {T}ransactions on {V}isualization and {C}omputer
  {G}raphics}, 16(6):973--979.

\bibitem[Wickham and Stryjewski, 2011]{wickham201140}
Wickham, H. and Stryjewski, L. (2011).
\newblock 40 years of boxplots.
\newblock Technical report.

\bibitem[Wilke, 2019]{wilke2019fundamentals}
Wilke, C.~O. (2019).
\newblock {\em Fundamentals of Data Visualization: A Primer on Making
  Informative and Compelling Figures}.
\newblock O'Reilly Media.

\bibitem[Wilkinson, 2012]{wilkinson2012grammar}
Wilkinson, L. (2012).
\newblock The grammar of graphics.
\newblock In {\em Handbook of Computational Statistics}. Springer.

\bibitem[Wright et~al., 2019]{wright2019primer}
Wright, T., Klein, M., and Wieczorek, J. (2019).
\newblock A primer on visualizations for comparing populations, including the
  issue of overlapping confidence intervals.
\newblock {\em The American Statistician}, 73(2):165--178.

\bibitem[Yau, 2024]{yau2024visualize}
Yau, N. (2024).
\newblock {\em Visualize This: The FlowingData Guide to Design, Visualization,
  and Statistics}.
\newblock John Wiley \& Sons.

\bibitem[Zgraggen et~al., 2018]{zgraggen2018investigating}
Zgraggen, E., Zhao, Z., Zeleznik, R., and Kraska, T. (2018).
\newblock Investigating the effect of the multiple comparisons problem in
  visual analysis.
\newblock In {\em Proceedings of the 2018 CHI Conference on Human Factors in
  Computing Systems}, pages 1--12.

\bibitem[Zhao et~al., 2017]{zhao2017controlling}
Zhao, Z., De~Stefani, L., Zgraggen, E., Binnig, C., Upfal, E., and Kraska, T.
  (2017).
\newblock Controlling false discoveries during interactive data exploration.
\newblock In {\em Proceedings of the 2017 ACM International Conference on
  Management of Data}, pages 527--540.

\bibitem[Zheng, 2017]{zheng2017data}
Zheng, J.~G. (2017).
\newblock Data visualization for business intelligence.
\newblock {\em Global Business Intelligence}, pages 67--82.

\end{thebibliography}

\newpage

\bigskip
\begin{center}
{\large\bf SUPPLEMENTARY MATERIAL}
\end{center}

Here we discuss details about course design for our data visualization courses in the Department of Statistics and Data Science at Carnegie Mellon University. Specifically, we outline the structure and content for our undergraduate and master's courses. Each course has its own intended audience and length. Thus, we hope these materials demonstrate how instructors in statistics-related fields can teach data visualization at the undergraduate and graduate level.

\subsection*{Undergraduate Course}

Our undergraduate course, titled Statistical Graphics and Visualization, is a 14-week course that serves approximately 100 students every fall and spring semester. As a prerequisite, students are required to have completed two semesters of introductory statistics, such that they have a basic understanding of hypothesis testing and statistical models. Our department and course are housed within Carnegie Mellon's Dietrich College of Humanities and Social Sciences; furthermore, the course counts towards the college's general education ``design'' requirement, which only needs to be completed by the time a student graduates. Thus, our course serves a wide range of majors, including economics, English, history, information systems, psychology, social and decision sciences, statistics, and other fields.

Some readers may wonder how the set of students can be so diverse when the prerequisite is seemingly strict. One of our college's general education requirements is that students must complete an introductory statistics course during their freshman year. Thus, students only have to take one additional statistics course in order to enroll in our course. Our reasoning for this prerequisite is to ensure that students have some familiarity with statistical inference, such that they can appreciate how inferential tools can complement data visualizations, which is one focus of the course.

Each week consists of two 50-minute lectures (one on Monday and one on Wednesday) and a 50-minute computer lab on Friday. Typically, 1-2 lectures are dedicated to ways to visualize particular data types (e.g., two-dimensional categorical data, one-dimensional quantitative data), and then 1-2 lectures are dedicated to inferential tools that can be used for that data type (e.g., chi-squared tests, Kolmogorov-Smirnov tests). Meanwhile, in the computer labs, students work through practice problems related to that week's lecture content; along the way, they can ask each other and the teaching staff questions, such that the computer labs act as an active learning component. The typical week-by-week timeline and set of topics for the course is shown in Table \ref{tab:undergradTimeline}. The first half of the semester focuses on visualizations and inferential tools involving one or two categorical and/or quantitative variables; this also corresponds to the taxonomy illustrated in Table \ref{tab:taxonomy}. Meanwhile, the second half of the semester focuses on topics involving more complex data types: high-dimensional, spatial, longitudinal, and text.

\begin{table}
    \centering
    \begin{tabular}{l|l}
    \hline
{\bf Week} & {\bf Topics}\\
\hline
Week 1  & Principles for Graphics, Data Structures, 1D Categorical Data\\
Week 2  & Inference for 1D Categorical Data \\
Week 3 & 2D Categorical Data \\
Week 4 & 1D Quantitative Data, Kernel Smoothing \\
Week 5 & Density Estimation, Inference for 1D Quantitative Data, Power \\
Week 6 & Scatterplots and Linear Regression \\
Week 7  & Nonlinear Regression \\
\hline
Week 8 & Pairs Plots, Contour Plots, Heat Maps, Distances \\
Week 9 & Clusters and Dimension Reduction: Dendrograms and PCA \\
Week 10 & Maps and Spatial Data \\
Week 11 & Time Series and Longitudinal Data \\
Week 12 & Text Data and Word Clouds\\
Week 13 & Text Data, Sentiment Analysis, Topic Modeling \\
Week 14 & Interactive Graphics \\
\hline
\end{tabular}
    \caption{Typical timeline and set of topics for our 14-week undergraduate data visualization course.}
    \label{tab:undergradTimeline}
\end{table}

As for assessments, we have students submit weekly homework assignments, which involve creating visualizations and conducting analyses with real datasets based on the past week's lecture and computer lab content. Furthermore, as a small participation grade, students are asked to critique a data visualization they found online or elsewhere once a month, in order to see how principles we discuss in class apply to real-world graphics. At midsemester, students complete a take-home midterm exam that assesses their ability to implement, navigate, and interpret the visualizations and inferential tools in Table \ref{tab:taxonomy}. After the midterm exam, students are organized into teams of 3-4 people, and by the end of the semester complete a final project that consists of visualizations and analyses of a real dataset that they choose. To ensure that students choose datasets that could lead to meaningful projects, we ask them to propose two possible datasets and explain how they fulfill criteria for the project (such as having enough variables to motivate at least 10 interesting graphs, a data structure complex enough to motivate at least one of the advanced topics in the second half of the semester, and an application nuanced enough to motivate at least three research questions). For their final project, teams must give a short oral presentation, submit a written report, and create an HTML document that is posted on our department website at \href{https://www.stat.cmu.edu/capstoneresearch/}{stat.cmu.edu/capstoneresearch/}. In this way, students not only demonstrate their ability to communicate statistical visualizations and analyses in several contexts, but also leave the class with a public-facing document that they can reference. Indeed, we encourage students to link their projects on their resumes, such that they can showcase their data science portfolio to future employers and graduate programs. We also use these online documents to run a graphics contest, where faculty and PhD students vote on the best graphics of the semester, such that undergraduates can get external validation of their work.

Finally, in terms of software, we have students use R to implement graphs and analyses (including the ggplot2 package), and RMarkdown to create their homework assignments and projects involving visualizations and discussions. We do not require students to have previous programming experience, and thus in some lectures and asynchronous material we give guidance on how to implement procedures in R. Furthermore, students gain further practice and guidance implementing visualizations and analyses in the weekly computer labs before completing homework assignments. That said, the course is not simply an R and ggplot2 tutorial, and we believe that the content and structure of the course could be readily applied to other programming languages or software.

\subsection*{Master's Course}

Our master's level course, titled Data Visualization, is an accelerated 7-week course that serves approximately 30 students in the first half of the fall semester. This course appears at the beginning of our Master of Science in Applied Data Science (MADS) curriculum, which spans two semesters and is targeted for students pursuing an industry position as a data scientist, data analyst, or data engineer. While there is similar content to the undergraduate course, there is a greater emphasis on computing and design, along with more advanced topics such as nonlinear dimension reduction techniques. An example of the course calendar and content is available at: \href{https://ryurko.github.io/mads-36613-fall24/}{\color{blue}https://ryurko.github.io/mads-36613-fall24/}. There, one can see how the first seven weeks of content in Table \ref{tab:undergradTimeline} is done in approximately two weeks, such that more time is spent on more advanced visualizations related to high-dimensional data, spatiotemporal data, text data, and interactive graphics with Shiny in R, as is appropriate for a graduate-level course. The website also includes the course's lecture slides and coding demos for implementing visualizations and analyses in R, which may be helpful resources for instructors wanting to develop their own data visualization course.

\subsection*{Data Availability Statement}

The data that support the findings of this study are openly available on the Open Science Framework (OSF) at \href{https://osf.io/wdehn/?view_only=81a99ddcf51f492393ea567b3d5f37dc}{https://osf.io/wdehn/?view\_only=81a99ddcf51f492393ea567b3d5f37dc}. They are also available in our replication materials at \href{https://github.com/ryurko/teaching-data-viz}{https://github.com/ryurko/teaching-data-viz}.

\end{document}